\documentclass[twocolumn,pre,superscriptaddress]{revtex4}
\usepackage{graphicx}
\newcommand{\ra}{\langle r \rangle}
\begin{document}
\title{Algorithms for 3D rigidity analysis and a first order
percolation transition}

\author{M. V.~Chubynsky}
\email{mykyta.chubynsky@umontreal.ca}
\affiliation{
D\'{e}partement de Physique, Universit\'{e} de Montr\'{e}al,\\
C.P. 6128, Succ. Centre-Ville, Montr\'{e}al, Qu\'{e}bec H3C 3J7, Canada}
\author{M. F.~Thorpe}
\email{mft@asu.edu}
\affiliation{
Department of Physics, Arizona State University,\\
P.O. Box 871504, Tempe, Arizona, 85287-1504
}

\date{\today}

\begin{abstract}
A fast computer algorithm, the pebble game, has been used successfully to
analyze the rigidity of 2D elastic networks, as well as of a special class of 3D networks,
the bond-bending networks, and enabled significant progress in studies of
rigidity percolation on such networks. Application of
the pebble game approach to general 3D networks has been hindered by the fact
that the underlying mathematical theory is, strictly speaking, invalid in this
case. We construct an approximate pebble game algorithm for general 3D networks,
as well as a slower but exact algorithm, the relaxation
algorithm, that we use for testing the new pebble game.
Based on the results of these tests and additional considerations, we argue that
in the particular case of randomly diluted central-force networks on BCC
and FCC lattices, the pebble game is essentially exact. Using the pebble game,
we observe an extremely sharp jump in the
largest rigid cluster size in bond-diluted central-force networks in 3D, with
the percolating cluster
appearing and taking up most of the network after a single bond addition. This
strongly suggests a first order rigidity percolation transition, which is in contrast to
the second order transitions found previously for the 2D central-force and 3D
bond-bending networks. While a first order rigidity transition has been observed
for Bethe lattices and networks with ``chemical order'', this is the first time
it has been seen for a regular randomly diluted network. In the case of site
dilution, the transition is also first order for BCC,
but results for FCC suggest a second order transition. Even in bond-diluted lattices,
while the transition appears massively first order in the order parameter (the
percolating cluster size), it is continuous in the elastic moduli. This, and the apparent non-universality,
make this phase transition highly unusual.
\end{abstract}


\maketitle

\section{Introduction}
\label{intro}

In this paper, we examine some important issues involved in studies of rigidity
of elastic networks in three dimensions. In general, the 3D rigidity problem is
much more difficult computationally than its 2D analog that has been studied
extensively using an efficient topological algorithm, the pebble game. In a way,
this parallels the Ising model that is much used for the theory of phase
transitions, in that exact solutions are available in 2D but not in
3D~\cite{baxter}. There is an important difference, however, that in rigidity,
exact means an algorithmic solution, rather than an analytic one.

We start by giving an extended discussion of the issues involved in 3D
rigidity that make it such a challenging problem. We then develop an exact
algorithm for small systems (up to several hundred or perhaps thousand sites)
where a solution for geometric quantities, such as identifying
the rigid clusters and the stressed regions, as well as counting the number of
floppy  modes, can be obtained. This serves as a
benchmark for a version of the much faster pebble game algorithm that we
develop, and shows that while the latter is approximate in general, for
central-force randomly diluted lattices its errors are very
small indeed and so that it may be regarded as
operationally exact. We then use the pebble game to study bond and site rigidity
percolation on FCC and BCC lattices. The transition appears to be massively
first order in all these cases, except for site percolation on the FCC lattice
where it is likely second order.

We hope our critical
assessment of this important problem will be useful for
mathematicians and computer scientists who are trying to develop rigidity
algorithms in 3D, for physicists interested in phase transitions, and for those
involved in studying the rigidity of 3D systems in the laboratory.

\subsection{Rigidity theory}
An elastic network is a set of {\it sites} in space and pairwise {\it constraints} between
these sites; a constraint is a statement that the distance between a certain
pair of sites has to take on a certain value, and if this value is different,
there is an associated potential energy cost. Constraints can be modeled as
elastic springs, although whether these springs are harmonic or not is
not important. Both static and dynamic properties of such an elastic network
depend in principle on the details of the potential energy function. However,
there are
certain properties that only depend on the geometry of the network. Among them:
the number of {\it floppy modes} (linearly independent infinitesimal motions
that do not deform any constraints
and thus do not cost energy); rigid cluster decomposition (a rigid cluster is
a set of sites that move together as a rigid body in any floppy motion);
some aspects of stress (particularly, what constraints are stressed); and some other related properties.
In fact, typically such properties are determined solely
by the network {\it topology} (i.e., what sites are connected to what sites by
constraints): most networks with different geometries but the same topology have
the same number of floppy modes, same rigid clusters, etc., and only an
infinitesimal fraction of {\it non-generic} networks that are special in some
way (for instance, having some constraints that are parallel) may differ in this
respect from the rest.
{\it Rigidity theory}~\cite{servatius, graver, whiteleyphysbio} deals with
properties of this type. Below, we outline some aspects of rigidity theory that
will be useful to us, without detailed proof or explanation; for details, see
the above references.
We will assume that the networks we are dealing with are always
{\it generic}. Generically, there is also no difference between rigidity with
respect to infinitesimal and finite deformations, and we will assume that this
is the case as well.

It is, of course, always possible and straightforward in principle to analyze
rigidity properties of a network by constructing a particular realization of
the potential energy function consistent with the given topology and then,
assuming infinitesimal displacements and thus the harmonic approximation,
diagonalizing the dynamical matrix to find the zero-frequency (floppy) modes;
likewise, stressed constraints can be obtained by relaxing the network, etc. Such
methods are, however, relatively slow, often scaling (particularly in the
case of dynamical matrix diagonalization) as the cube of the network size; also,
they are subject to inevitable round-off errors, which may, for instance, make
zero eigenvalues of the dynamical matrix slightly non-zero etc. Yet methods of
this type are the only choice if one needs detailed information, such as the
actual values of the frequencies of all modes, the eigenvectors of the dynamical
matrix, etc. (assuming, of course, that the actual potential energy function,
and not just the topology, is known). But if we are only interested in those
properties that depend solely on the topology, clearly such methods are somewhat
``irrational'': they take some irrelevant information (all force
constants and detailed geometry) as input, and produce a lot of unnecessary
output (such as all the eigenvalues and eigenvectors of the dynamical matrix,
while we only want the number of zero eigenvalues). Thus there may be more
rational methods that deal with network topology directly, never constructing
a physical realization in the process. Such methods may be faster and also,
since the topology information is discrete, not subject to round-off errors.
Indeed, such methods have been found, as discussed below.

Historically, the first and simplest of such methods (but, unfortunately, only
approximate) is {\it Maxwell counting}~\cite{maxwell}, due to J.C.~Maxwell. Consider
a $d$-dimensional elastic network of $N$ sites. Dimensionality $d$ here (and elsewhere
in this paper) refers to the dimensionality of
space in which the sites can move, rather than the dimensionality of the network
itself. It is possible that some of the constraints in the network cannot be
satisfied simultaneously; it is useful, {\it just for the sake of this
consideration}, to change their lengths so that they fit exactly and are not
strained; it turns out that this does not affect the number of floppy modes in
generic networks and so can be used for the floppy mode counting that we do here
(but cannot be used to find stress, of course). If the network has no constraints,
all degrees of freedom correspond to floppy modes whose number is thus $dN$.
Each constraint, in the linear approximation, is some linear relation between
the coordinates of sites in the system. Then, when a constraint is added to the
network, it reduces the dimensionality of the space of allowed infinitesimal
motions by 1,
{\it if} the relation between the coordinates that this constraint
represents is linearly independent from the rest. If one assumes that all
constraints are like this (i.e., {\it independent}), then the number of floppy
modes is
\begin{equation}
F_{\rm Maxw}=dN-N_c,
\end{equation}
where $N_c$ is the number of constraints and $F_{\rm Maxw}$ denotes the number of
floppy modes in the Maxwell counting approximation. Of course, this result
for the number of floppy modes is only approximate,
since in reality not all constraints are independent. In a generic network,
non-independent constraints are those and only those that are
inserted between two sites that are already mutually rigid even before the
insertion --- such a constraint does not restrict floppy motions
further and in this sense is {\it redundant}; it does not decrease the
number of floppy modes. So the correct result for $F$ is actually
\begin{equation}
F=dN-N_c+N_R,\label{Maxexact}
\end{equation}
where $N_R$ is the number of redundant constraints. The Maxwell counting
approximation neglects $N_R$; a more accurate method would require a way to
evaluate it. Note that if the lengths of constraints are restored (i.e., they
are no longer required to fit exactly), then generically, a redundant constraint
inserted in the network becomes strained and introduces stress. Thus (again
generically) stress is present if and only if there is redundancy ($N_R\ne 0$).

Since the number of floppy modes in a network of size $N\ge d$ can never be less
than $d(d+1)/2$ (the number of motions of a rigid body), then it is
clear that if $dN-N_c<d(d+1)/2$ for the network {\it or} for any of its
subnetworks of size $N_s\ge d$, then $N_R>0$ (i.e., there must be some redundant
constraints). A more interesting question is if the converse is also true. In
other words, if $N_R>0$, is it true that there is at least one subnetwork of
size $N_s\ge d$, for which $dN_s-N_{cs}<d(d+1)/2$ ($N_{cs}$ being the number of
constraints within the subnetwork)? In 2D ($d=2$), for generic networks the
answer is yes, and this
statement is known as the {\it Laman theorem}~\cite{laman}. It is also
{\it conjectured} to
be true for a particular class of 3D networks, known as {\it bond-bending}
networks. A bond-bending network is defined by a set of {\it bonds}; constraints
then connect all first neighbors ({\it bond-stretching} or {\it central-force}
constraints) and all second neighbors ({\it bond-bending} or {\it angular}
constraints). The conjecture is a part of what is known as
the {\it molecular framework conjecture}~\cite{tay, whiteley, whiteleyphysbio};
the origin of this name is the fact
that bond-bending networks are a natural model for covalent molecules (as well
as covalent disordered solids), since covalent bonds have strong bond-stretching
and bond-bending interactions associated with them, and all other
interactions are usually weaker. This conjecture is not proved rigorously;
however, no counterexamples have been found in more than 20 years since its
formulation. For general (non-bond-bending) 3D networks there are violations of
the molecular framework conjecture, as discussed in detail below. Note that
for bond-bending networks, we are making a careful distinction between
{\it bonds} and {\it constraints}: the set of bonds specifies what sites are
considered first neighbors (those that are connected by a bond), and then
constraints connect both first and second neighbors. On the other hand,
in {\it central-force networks} that we consider in Section~\ref{CFnets},
there is one constraint per bond and we use ``bonds'' and ``constraints''
interchangeably.

The Laman theorem in 2D and the molecular framework conjecture for 3D
bond-bending networks enable a convenient and fast approach to finding $N_R$
exactly. Start with the ``empty'' network (all sites present but no
constraints). Such a network obviously
has $N_R=0$. Now, add constraints one by one checking each of them for
redundancy by testing all subnetworks that the newly added constraint belongs
to. If the constraint is redundant, $N_R$ is increased by one; otherwise it
is unchanged. Thus, $N_R$ is known at all times during the network construction
process. One very important caveat in the case of 3D bond-bending networks is
that even if the final network being analyzed is bond-bending, this
is not necessarily true for the intermediate networks obtained during the
construction process (and, in fact, cannot in general be true for all of them).
For this reason, it is important to keep the networks as close to being
bond-bending as possible. Namely, a constraint coinciding with a bond (a
first-neighbor constraint) should always be inserted first, and all
second-neighbor constraints induced by the just inserted first-neighbor
constraint (i.e., those second-neighbor constraints that span the angle formed
by the just inserted first-neighbor constraint with previously inserted
first-neighbor constraints) should follow immediately, before any other
first- or second-neighbor
constraints are inserted. While this does not keep the network strictly
bond-bending at all times, deviations are as small as possible,
and it is assumed (as a part of the molecular framework conjecture) that all intermediate
networks obey the statement of the conjecture as well.

Another issue is rigid cluster decomposition. Rigid clusters in 2D and in 3D
bond-bending networks have a useful property: they are
always {\it rigid by themselves}, i.e., they remain rigid when separated from
the rest of the network. A corollary of this is that rigid clusters in
such networks are always {\it contiguous}: when moving from any
site of the cluster to any other site belonging to it along the network
constraints, it is always possible to choose a path such that only sites
belonging to the same cluster are passed. Simply stated, rigid clusters always
``come in one piece.'' These properties mean that, first of all, a rigid cluster of size $n$ with all
redundant constraints removed will always have exactly $dn-d(d+1)/2$ constraints
within itself; also, contiguity allows easy cluster mapping, by starting with
an arbitrary group of $d$ mutually rigid sites and then moving outwards until the
region that has emerged is fully surrounded by sites not rigid with respect to
at least one of the $d$ initial sites.

Regarding stress determination, an important fact is that in 2D and 3D bond-bending
networks, the set of all stressed constraints can be represented as the
union of {\it stressed regions}, each of which is {\it stressed by itself},
i.e., remains stressed when separated from the rest of the network, and has a
property that it is possible to find a
set of sites such that all constraints connecting sites within the set belong
to the region and all constraints connecting sites at least one of which does
not belong to the set do not belong to the region. In other words, using graph
theory terminology, each stressed region is an {\it induced subgraph} of the graph whose
vertices are the network sites and whose edges are the constraints.

An algorithm using the above ideas, known as the {\it pebble game}, was
proposed first for 2D networks~\cite{jacobs95, jacobs97, site} and then for 3D
bond-bending networks~\cite{jacobs98, travbio, site}. The idea
is to relate the constraints to the degrees of freedom
for all subnetworks simultaneously, by assigning {\it pebbles} to
degrees of freedom and then matching those pebbles to constraints. The details
of the 3D version of the algorithm are described in the next subsection.

From the description of the pebble game, it will become clear that it relies
significantly on the three special properties of 2D and 3D bond-bending networks
stated above: the molecular framework conjecture; the contiguity of rigid
clusters and their being rigid by themselves; and stressed regions being induced
subgraphs. In Section~\ref{nonBB} we demonstrate that unfortunately, all of
these properties are violated in general in 3D networks that are not fully
bond-bending. Creation of a pebble-game-type algorithm that does not rely on
these properties is problematic and a way to do this has not been found to date.
While partial fixes (covering some but not all possible situations) are
possible, in our generalization of the pebble game that we introduce in
Section~\ref{peb}, we choose to ignore these
problems completely. This, of course, makes the algorithm only approximate, and
to estimate the accuracy of the new pebble game, an exact but slower algorithm, the
{\it relaxation algorithm}, is introduced in Section~\ref{relax}. Some general
considerations on the accuracy of the pebble game and how this accuracy can be
estimated using the relaxation algorithm are given in Section~\ref{errors}. In
the rest of the paper, we look at a few applications of the new algorithms. In
Section~\ref{CFnets}, we consider central-force networks (i.e., those that have
only first-neighbor constraints) obtained by randomly
removing bonds (bond-diluted networks) or sites (site-diluted networks) from
regular three-dimensional lattices. In this case,
we show that the pebble game is essentially exact for the most interesting
quantities, such as the number of floppy modes and the size of the largest  rigid
cluster, although some very small clusters may be misidentified. On the other
hand, the pebble game is much less successful in some other cases, as we show in
Section~\ref{edgesharing}. We use the success of the pebble game for randomly
diluted central-force networks to study {\it rigidity percolation} on such
networks. An introduction to the subject of rigidity percolation is given in the
last subsection of this section, after the description of the old pebble game
algorithm.

\subsection{The pebble game algorithm for bond-bending networks}
The 3D bond-bending version of the pebble game algorithm is as
follows~\cite{jacobs98, travbio, jacobspc}. Starting with the empty network without
constraints, three {\it pebbles} are assigned to each site, so that the total number
of pebbles is equal to the total number of degrees of freedom, $3N$. A pebble
can be free, or it can cover one of the constraints associated with the site
to which the pebble belongs. Initially, there are no constraints, so all pebbles
are free. As constraints are added to the network, all {\it independent}
(non-redundant) constraints (detected as described below) are covered by a
pebble from either side and should
remain covered at all times during the pebble game process. Since the number of
independent constraints is $N_c-N_R$, then, according to Eq.~(\ref{Maxexact}),
the number of free pebbles is equal to the number of floppy modes. A
constraint can be covered at either end, and this allows freeing of pebbles.
A pebble covering a constraint can be freed, if there is a free pebble
available at the other end of the constraint; then that free pebble covers the
constraint and the pebble covering it is released; this process may have to be
repeated several times, if a pebble is free not at the end of the constraint,
but at one of its neighbors, neighbors of neighbors etc. As a consequence,
checking for whether freeing a pebble at a given site is possible starts at that
site, then looks at what constraints the pebbles belonging to the site cover and
checks the site's neighbors connected to it by those constraints; if no free pebbles
are found there, the procedure is repeated until a free pebble is found or
until no unchecked sites connected to checked sites by constraints covered by pebbles
belonging to checked sites remain. If the search for a pebble has failed, the
region over which the search has proceeded (the failed pebble search region) is
recorded, which is important for finding stressed regions, as described below.

Each newly
inserted constraint is tested for independence in the following way. First,
all three pebbles need to be freed at each of the two ends of the new
constraint --- this is always possible, unless the new constraint coincides with
a previously inserted one, in which case the new constraint is obviously
redundant and should not be tested. Then, with all six pebbles kept free,
an attempt is made to free one more pebble at each neighbor of the ends of the
new constraint in turn. In fact, even fewer checks
are needed: for a bond-stretching (first-neighbor) constraint, just first
neighbors at one end need to be checked (second neighbors need not be
checked, even though they are connected with a second-neighbor constraint);
for a bond-bending (second-neighbor) constraint, only the vertex of the angle
that the constraint spans needs to be checked.
If {\it all} of these attempts are successful, then the
new constraint is independent and should be covered by one of the six pebbles
at its ends. Otherwise, the constraint is redundant and should not be covered.
As a reminder, constraints should be inserted in a
particular order: a first-neighbor constraint is inserted first and then all
second-neighbor constraints induced by it should be inserted immediately
afterwards before another first-neighbor constraint is inserted. Just as 
the very
similar 2D algorithm relies on the Laman's theorem, as explained in detail in
Ref.~\cite{jacobs97}, the above procedure assumes the validity of its
generalization, the molecular framework conjecture.

Whenever a redundant constraint is inserted, it will create additional stress
in the network and may increase the set of constraints that are stressed.
The redundancy is detected when a pebble search fails, and the region of the
failed pebble search should be merged with such regions found previously to
find the part of the network that is stressed. Once one failure to
find the pebble is detected, there is no need to continue further checks in
order to find the stressed region --- the failed search regions will coincide
for all neighbors for which the search fails. Regions of failed pebble search
are defined as sets of {\it sites} passed when searching for a pebble, and
{\it all} constraints connecting such sites are stressed. Of course, this
implies an important property of bond-bending networks that we have already
mentioned --- that any stressed region is always an induced subgraph of the graph
whose vertices are the sites and whose edges are the constraints.

During the pebble game, whenever a large stressed region is detected, the
{\it tetrahedralization} procedure~\cite{jacobspc} (similar to the
triangularization procedure in 2D~\cite{jacobs97, jacobs96}) is commonly
done to convert the stressed
region into an isostatic (rigid but unstressed) one. This speeds up further
pebble searches significantly. We do not consider this procedure here and do
not implement it in our treatment of general (non-bond-bending) 3D networks.
This limits the network sizes we can routinely consider to perhaps $10^5$
sites or so, while with tetrahedralization $10^6$-site or even larger
networks could be considered, so the implementation of this procedure in the
future is desirable.

Once the network construction is finished, the information on the number of
floppy modes and the stressed constraints is available. The next stage is rigid
cluster decomposition. One thing to keep in mind is that unlike in the usual
{\it connectivity} percolation, a site can belong to several clusters
simultaneously (imagine, for instance, two rigid objects sharing a common point,
a {\it pivot}, or a common axis, a {\it hinge}); but choosing three mutually
rigid sites identifies a cluster uniquely: any three chosen sites can
belong simultaneously to at most one cluster. Bond-bending networks are special:
among all clusters to which a given site belongs, there is always one and only
one to which all of its neighbors also belong. For this reason, for a
bond-bending network rigid cluster decomposition can be given by specifying
for each site the unique cluster to which this site belongs with all of its
neighbors. Given the above, it is
convenient to start mapping a rigid cluster by choosing a site
and two of its first neighbors. Such three sites are always mutually rigid
(indeed, they form an angle and bond angles are constrained in
bond-bending networks) and
thus specify a rigid cluster. A maximum number of pebbles are freed at the
three chosen sites. It is always possible to free exactly
6 pebbles.
After this, neighbors of the
three chosen sites are picked in turn, and for each of such sites an attempt
is made to free a pebble while keeping the six freed pebbles at the first three
sites free. If freeing an extra pebble fails, the site being tested is rigid with
respect to the first three; moreover, this is true for the whole
region of the failed search. But if freeing an
extra pebble succeeds, then the site is not rigid with respect to the first
three sites. After
all neighbors of the three initial sites are checked and if at least some are
found rigid with respect to the initial sites, neighbors of the neighbors
found rigid are checked etc. The process continues until no unchecked neighbors
of the sites deemed rigid with respect to the initial three sites remain. At
this point mapping of
the cluster is complete, since all clusters are contiguous. All sites such that
all neighbors of them are found
rigid with respect to the original three sites are assigned identical labels to
specify the cluster. Then another site is chosen among those that are not yet
labeled, together with its two neighbors, and mapping another cluster starts.
This continues until all sites are labeled, at which point rigid cluster
decomposition is complete. Note that this procedure uses the contiguity of
rigid clusters, but also, less explicitly, their being rigid by themselves,
since a region of the network not rigid by itself may contain more than six
free pebbles, and thus extra pebbles besides the six freed at the first three
sites may be found.

After the rigid cluster decomposition process described above, for each bond in
the network its end sites can have either identical or different labels. In the
latter case, both ends of the bond are shared between two rigid clusters and
such a bond is a {\it hinge}. It is impossible for two clusters in a
bond-bending network to share just a single site, so {\it pivots} cannot exist
in such networks~\cite{jacobs98}. Only first-neighbor constraints coinciding with
bonds can be hinges; this can never happen to second-neighbor constraints. It
is also impossible to have a hinge that does not coincide with any constraint
(such hinges are known as {\it implied}). A hinge can be shared between at most
two clusters. These limitations do not apply to non-bond-bending networks.

To avoid confusion, we should note that there is also a different variant of
the 3D pebble game for bond-bending networks based on an equivalent
representation of such networks as {\it body-bar}
networks~\cite{jacobs98, travthorpe}. It is this variant
that is used, for instance, in the FIRST software for protein rigidity
analysis~\cite{proteins01, pnas02, virus, site}. However, as the body-bar
representation does not apply to
non-bond-bending networks, the corresponding variant of the pebble game
algorithm does not extend naturally to such networks, and we do not consider
it here.

\subsection{Rigidity percolation}
\begin{figure}
\begin{center}
\includegraphics[width=3in]{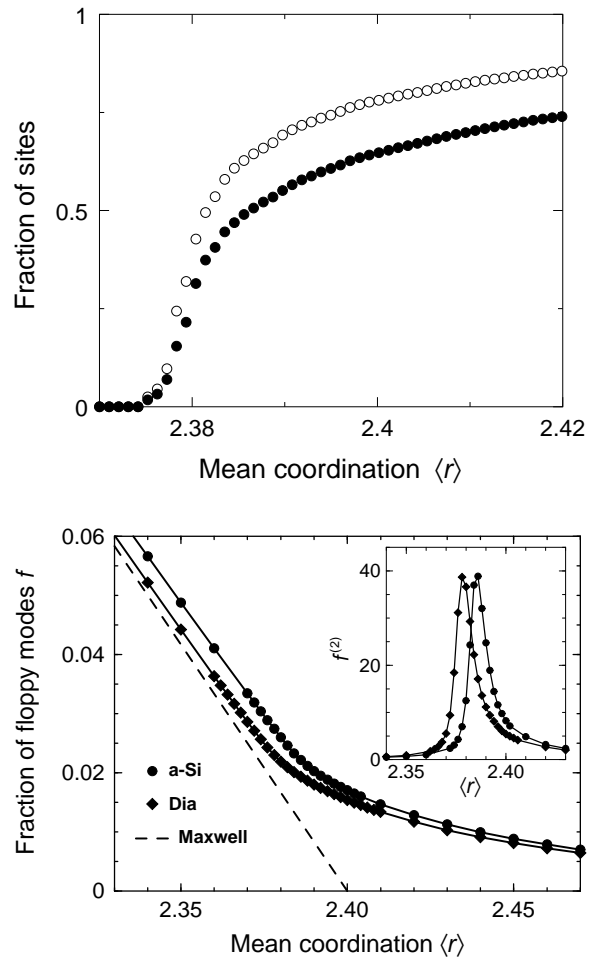}
\end{center}
\caption{ Bond-bending networks in 3D. (Top) The fractions of sites belonging to the percolating rigid
cluster (open circles) and the percolating stressed region (filled circles)
as functions of mean coordination $\ra$ for the case of random bond dilution of
the diamond lattice. The results are averages over 11 networks of 125 000 sites
each. Rounding near the transition is due to finite-size effects. (Bottom)
The number of floppy modes per degree of freedom $f=F/3N$ for randomly bond
diluted amorphous silicon (circles) and diamond lattices (diamonds). The dashed line is the
Maxwell counting result. The inset shows the second derivative of $f$ with
respect to $\ra$. The upper panel is from Ref.~\cite{chemorder}; the lower
panel is adapted from Ref.~\cite{travthorpe}.}
\label{perc3DBB}
\end{figure}

\begin{figure}
\begin{center}
\includegraphics[width=2.8in]{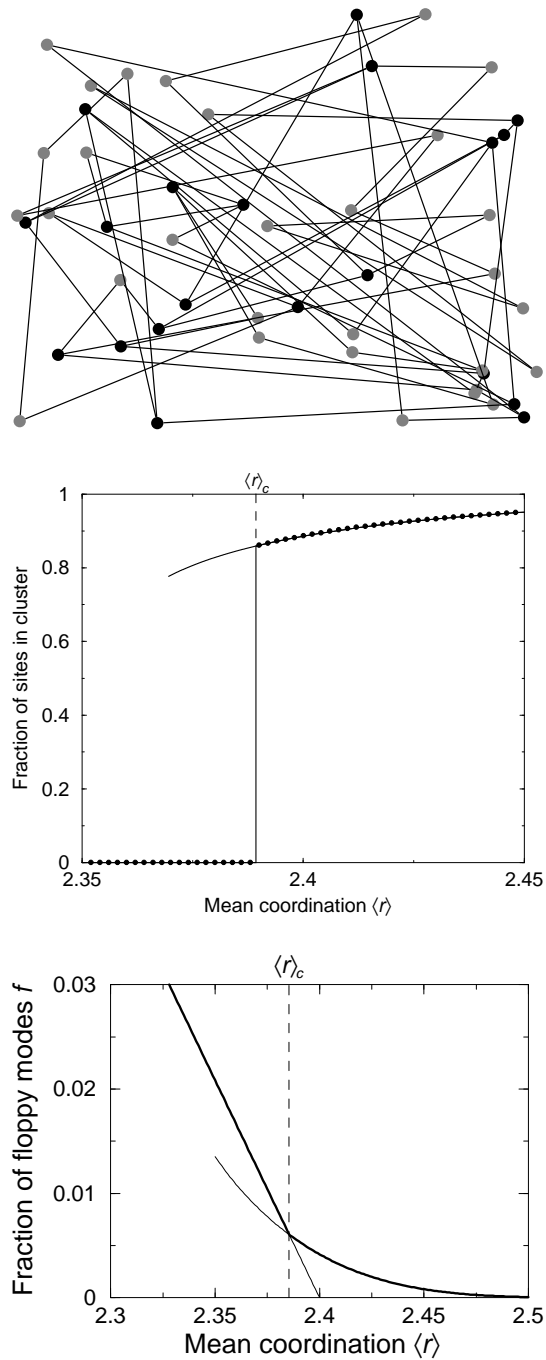}
\end{center}
\caption{(Top) A sketch of a random bond network: sites are connected
at random, regardless of the distances. This network consists of 2- and
3-coordinated sites (gray and black, respectively). (Middle) The fraction of
sites in the percolating rigid cluster
as a function of $\ra$ for a bond-bending random bond network consisting of 2-
and 3-coordinated sites in 3D. The solid line is theoretical, the circles are
the result of pebble game simulations. The transition occurs at
$\ra_c$. (Bottom) The theoretical number of floppy modes per degree of
freedom, $f=F/3N$, for a bond-bending random bond network
consisting of 2- and 3-coordinated sites in 3D.
Note the break in the slope at the transition. These panels are adapted from
Refs.~\cite{travthorpe} and \cite{thesis}.}
\label{bethe}
\end{figure}

The concept of rigidity percolation was first introduced by
Thorpe~\cite{thorpe83} in the
context of covalent network glasses; it was subsequently studied soon after
in more detail for central force networks by Feng and Sen~\cite{fengsen} and by
Feng, Thorpe, and Garboczi~\cite{garboczi}. In network glasses, covalent bonds
connecting atoms are strongly directional, meaning that there is a strong restoring
force associated with changing both bond lengths and bond angles. At the same
time, all other interactions are much weaker. For this reason, covalent glasses
can be modeled as 3D bond-bending networks, for which the pebble game is exact.
If the number of bonds per site is increased (in practice, by changing the
chemical composition), the networks go from overall floppy to overall rigid as
the rigidity percolation threshold is crossed, at which point a percolating
rigid cluster emerges. The fraction of sites in the percolating cluster (that
serves as the order parameter for the rigidity percolation transition) grows
continuously, starting from zero at the threshold, and thus the transition is
said to be continuous, or
{\it second order}~\cite{travthorpe, czech01, currop} (Fig.~\ref{perc3DBB}). One can also
look at {\it stress percolation} --- whether the set of bonds that are stressed
percolates. The stress percolation in random networks occurs at the same point as rigidity
percolation and the behavior is similar (Fig.~\ref{perc3DBB}), although there
are models in which network self-organization leads to these thresholds being
different~\cite{thorpe00, czech01, currop, micoulaut02, micoulaut03, barre, chubynsky06}. Likewise,
rigidity percolation
in 2D {\it central-force} networks was considered (2D bond-bending percolation
is equivalent to the usual, connectivity percolation). The result is also a second order
transition~\cite{jacobs95, jacobs96}. But there are known cases in which the
transition is {\it first order}, i.e., the fraction of sites in the percolating
cluster {\it jumps} from zero to a non-zero value at the transition. This has
been found for ``pathological'' Bethe lattices or random bond
networks~\cite{dux97, dux99, travthorpe} (Fig.~\ref{bethe}) and networks with chemical
order~\cite{chemorder}, but up to now, there have been no cases where it would
be observed for a regular randomly diluted network.

Besides the size of the percolating cluster, the order of the rigidity
percolation transition can be found by looking at the behavior of the number of
floppy modes $F$ as a function of {\it mean coordination} $\ra$ (the
average number of bonds connecting a site to other sites). It has been
suggested~\cite{dux99} that $-F$ serves as an analog
of the free energy of the system. When a system goes through a phase transition,
the free energy is continuous, but in a first order transition, its first
derivative is discontinuous; in a second order transition, the first derivative
is continuous, but the second derivative shows a singularity. Indeed, in the
case of a regular 3D bond-bending network (Fig.~\ref{perc3DBB}), the number of
floppy modes is a continuous and smooth function of $\ra$, but there is a
cusp in the second derivative, which is consistent with a second order
transition; on the other hand, for the random bond network, 
where the transition is first order, there is a break in the first
derivative (see the lower panel of Fig.~\ref{bethe}).

There have also been some studies of 3D central-force elastic
networks~\cite{garboczi, garbsite, sahimi}. However, networks used by Feng
{\it et al.}~\cite{garboczi} and Garboczi and Thorpe~\cite{garbsite}
were not large enough to make definite conclusions about the nature of the
rigidity transition. While Arbabi and Sahimi~\cite{sahimi} used
larger networks, they only
considered physical properties like elastic moduli and force distributions; they
were not concerned with geometric quantities, such as sizes of rigid clusters.
Lack of a fast pebble game-type algorithm made studies of 3D central-force
rigidity extremely difficult. Since it was known that the pebble game
is not exact for 3D networks that are not bond-bending, it was assumed that the
errors would make any applications of the pebble game approach unreliable. In
Section~\ref{CFnets}, we show that this is not the case and then use the pebble
game to study the rigidity percolation transition in both the bond-diluted and
the site-diluted networks in 3D.

\section{Non-bond-bending 3D networks}
\label{nonBB}
For non-bond-bending networks in 3D, unfortunately, the
molecular framework conjecture and other statements crucial for the application
of pebble-game-type algorithms are not true in general. In this section we
present a few known counterexamples. Some of these were published before
(see, e.g., Ref.~\cite{jacobs98}).

\begin{figure}
\begin{center}
\includegraphics[width=3in]{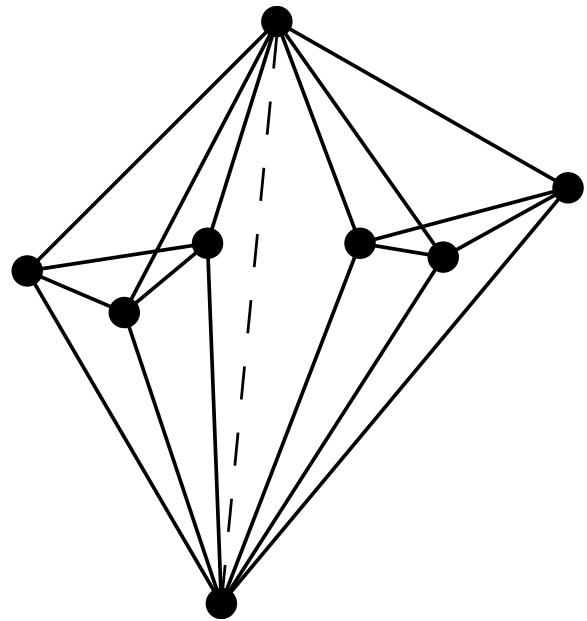}
\end{center}
\caption{An example of a network (the double-banana graph) for which the
generalization of the Laman theorem fails. The dashed line is a
hinge around which the two ``bananas'' can rotate.}
\label{2banana}
\end{figure}

Figure~\ref{2banana} shows an example of a network for which the generalization
of the Laman theorem fails. This is an infamous
{\it two-banana graph}~\cite{servatius}.
For all subnetworks with $N_s\ge d=3$,
$dN_s-N_{cs}\ge d(d+1)/2$ and so there should be no redundant constraints,
$N_R=0$. Since there are $N=8$ sites and $N_c=18$ constraints, there should be
$F=3\times 8-18=6$ floppy modes --- exactly the number that a rigid body has
(3 translations and 3 rotations), so the network should be rigid. It is obvious
that this is not the case, as the two ``bananas'' can rotate around the
axis they share. Thus, there is one internal
floppy mode in addition to the 6 rigid body motions, so $F=7$ and then,
according to Eq.~(\ref{Maxexact}), $N_R=1$ --- there is one redundant
constraint.

\begin{figure}
\begin{center}
\includegraphics[width=3in]{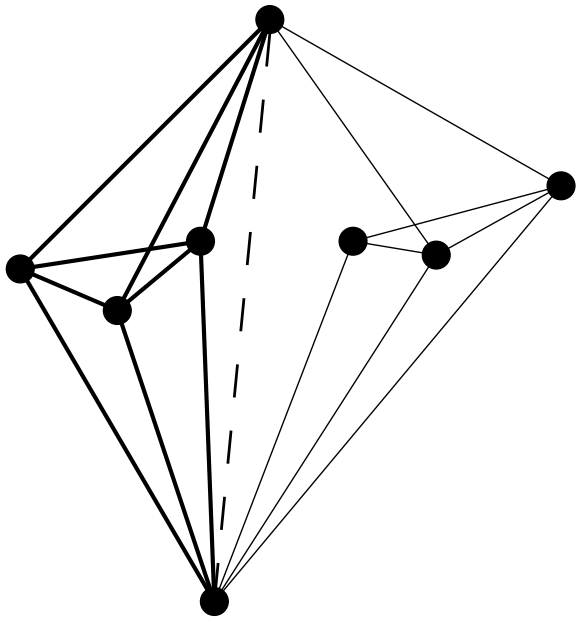}
\end{center}
\caption{An example of a network with a rigid cluster (shown with thinner lines)
that is no longer rigid when taken in isolation from the rest of the network.
The dashed line is a hinge.}
\label{2banana1}
\end{figure}

\begin{figure}
\begin{center}
\includegraphics[width=3in]{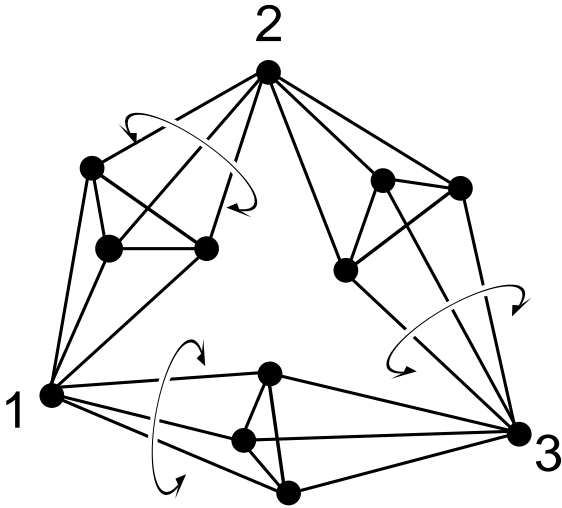}
\end{center}
\caption{An example of a network with a non-contiguous rigid cluster consisting
of sites marked 1, 2, 3.}
\label{3banana}
\end{figure}

Likewise, rigid clusters are no longer necessarily rigid by themselves or even
contiguous. Figure~\ref{2banana1} shows the same network as in Fig.~\ref{2banana},
except one constraint is missing. There is still one internal floppy mode, as in
Fig.~\ref{2banana}, but
as $N_c$ is less by one, $N_R=0$, which agrees with the generalization of the Laman
theorem. Yet, the part of the network shown with thin lines is a rigid
cluster, despite not being rigid by itself, when ``detached'' from the rest of
the network that rigidifies it.
A straightforward application of the rigid cluster decomposition procedure
described above may fail to detect this rigid cluster. Figure~\ref{3banana}
shows an even more extreme example of a {\it non-contiguous} rigid
cluster~\cite{jacobs98}. The three
``bananas'' in the figure are ``normal'', contiguous rigid clusters. But in
addition to that, sites marked 1, 2 and 3 also form a rigid cluster, being
mutually rigid with no other sites in the network rigid with respect to all
three. This cluster is, of course, non-contiguous, and there is no way a
rigid cluster decomposition procedure similar to the one described above can
detect it, as it marks clusters in a contiguous fashion.

There is one common feature in the networks shown in
Figs.~\ref{2banana}--\ref{3banana}. In all three
cases, there are {\it implied hinges} (shown in the first two figures with dashed lines).
As a reminder, a hinge is a straight line that for rigid clusters sharing
exactly two sites goes through these two sites; it is the axis of rotation
around which the clusters can rotate with respect to each other. In
bond-bending networks, all hinges coincide with constraints (i.e., are
{\it explicit}), as mentioned above, but in general, this is not necessarily
the case, as Figs.~\ref{2banana}--\ref{3banana} illustrate. In
Figs.~\ref{2banana} and \ref{2banana1}, there are two rigid
clusters, and they share an implied hinge. In Fig.~\ref{3banana}, there are
three implied hinges, each shared by the non-contiguous cluster 1--2--3 with one
of the three ``bananas''.

In fact, it turns out that problems with obeying the generalization of the
Laman theorem (or the molecular framework conjecture), as
well as with contiguity of rigid clusters and their being rigid
by themselves are {\it always} due to implied hinges. A network not having
implied hinges has no such problems, and if implied hinges are placed
explicitly as constraints, the problems are eliminated as well. This can be
checked explicitly for networks in Figs.~\ref{2banana}--\ref{3banana}. If the
hinge is placed explicitly as a constraint in Fig.~\ref{2banana}, there are
now $N_c=19$ constraints, and condition
$dN-N_c\ge d(d+1)/2$ is now violated, so the redundant constraint is now
predicted correctly to exist. When the hinge is placed explicitly in
Fig.~\ref{2banana1}, it
becomes part of the rigid cluster drawn with thin lines, and this cluster then
becomes rigid by itself. Finally, in Fig.~\ref{3banana}, once the hinges are placed
explicitly, the rigid cluster 1-2-3 becomes contiguous. The general statement
that all problems with floppy mode counting and rigid cluster decomposition are
due to implied hinges is related to the so-called
{\it Dress conjecture}~\cite{tibor} in
rigidity theory. The Dress conjecture actually gives the exact count of the
number of floppy modes once all implied hinges are identified. But since the
implied hinges still need to be found first, unfortunately, unlike the molecular
framework conjecture,
the Dress conjecture does not give rise to a straightforward approach to
floppy mode counting or finding rigid clusters, and at this time, there is no
topological algorithm of the
pebble game type that would do that. Of course, there can be much more
complicated cases than those shown in Figs.~\ref{2banana}--\ref{3banana} ---
whole hierarchies of bananas within bananas within bananas ---
and a way to take all of such cases into account has not been found to date.

\begin{figure}
\begin{center}
\includegraphics[width=3in]{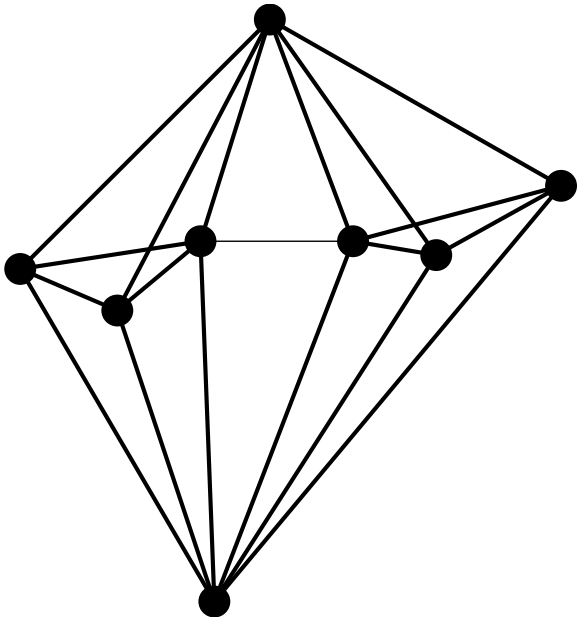}
\end{center}
\caption{An example of a network in which the
stressed region is not an induced subgraph, since the thin constraint is the only
one that is not stressed.}
\label{bananalocked}
\end{figure}

There is also a complication related to stress determination. The pebble game
finds stressed regions as sets of sites such that
all constraints connecting
sites within the same set are stressed and all the rest are unstressed. As
explained above, this involves an implicit assumption that all
stressed regions are induced subgraphs. But in
non-bond-bending networks it need not be so. Consider the network
consisting of two bananas with one bridging
constraint between them, as in Fig.~\ref{bananalocked}. All constraints are
stressed, with the exception of the ``bridge'' (a thinner line in the figure)
Note that any subset of the set of stressed constraints (other than the
full set) cannot be considered a stressed region, because it would not be
stressed in isolation. For instance, when the two bananas are separated from
each other, each banana separately will not remain stressed. So the whole set
of stressed constraints is a single stressed region here. But this stressed
region is not an induced subgraph: it is impossible to find a set of sites such
that all constraints but the ``bridge'' connect the sites in the set, but at
least one of the ``bridge'' ends is outside the set. A pebble game procedure based on
failed pebble search regions will not be able to identify this
stressed region correctly. Note
that in this case, there are no implied hinges. So whereas implied hinges are
the reason for all problems with floppy mode counting and rigid cluster
decomposition, this is not so for stress. But note also that if the
``problematic'' ``bridge'' constraint is removed, the hinge {\it will} appear.
It can be argued that whenever there are problems in the pebble game
determination of stress, they are either due to implied hinges (or
uncovered explicit hinges --- see Section~\ref{errors}), or such a
hinge would appear if one constraint is removed.

\section{The pebble game for non-bond-bending networks}
\label{peb}

The lack of an exact pebble-game-type algorithm for general 3D networks has
been a significant impediment to studying such networks. In Section~\ref{intro},
we have seen that there are many properties of bond-bending networks
that are useful for the pebble game algorithm and
that do not hold in general for networks that are not bond-bending. Violation
of some of these properties, such as the possibility to specify rigid cluster
decomposition by labeling sites, is a mere inconvenience. Crucial, however, is
the absence of implied hinges in bond-bending networks and their presence in
general non-bond-bending networks, and as a consequence, the violation of the
molecular framework conjecture, of the properties of contiguity of rigid
clusters and their being rigid by themselves, as well as the induced subgraph property
of stressed regions. These properties are essential for the pebble game approach
and it is not known how to avoid using them in a pebble-game-type algorithm.

But even though we
know that in some cases application of the pebble game approach
would be wrong, a reasonable question to ask is just how wrong such
an algorithm would be in various cases of interest. In other words, are the errors
frequent and significant or are they rare and negligible? In much of the
remainder of the paper, we will try to answer these questions. In this
section, we describe the pebble game algorithm we are going to use, which is
mostly a straightforward generalization of the algorithm for bond-bending
networks described above, except that it does not rely on certain 
properties of bond-bending networks that no longer hold for non-bond-bending
ones. This algorithm needs to be compared with the exact
result, and for this reason in the next section we introduce a
``physics-based'' approach, the {\it relaxation algorithm}. It has all the
disadvantages mentioned before, such as slowness and round-off errors (although
it is likely faster than straightforward diagonalization; see the discussion at
the end of Section~\ref{relax}) --- but if
the latter are brought under control, the approach is potentially exact and
can be used for testing the pebble game.

The first part of the pebble game algorithm, in which the redundant constraints
are counted and stress is detected, is very similar to the
bond-bending case, but some details differ. Just as for bond-bending networks,
a constraint is tested for independence by first freeing six pebbles at its ends
and then attempting to free an extra pebble at the neighbors of an end of the
constraint in turn. Obviously, since the subdivision
into first- and second-neighbor constraints is no longer present, a particular
order in which constraints are inserted can no longer be enforced --- although
for networks with partially bond-bending character (for instance, a
bond-bending network with some angular constraints missing), following the
order (a first-neighbor constraint inserted first and all associated
second-neighbor constraints immediately afterwards) as much as possible could
be beneficial and reduce errors (but see a counterexample to this in
section~\ref{edgesharing}). Another difference is that, again, for the
same reason that there is no longer a strict subdivision into different types of
constraints, {\it all} neighbors of at least one of the ends of the constraint being
tested should be checked by trying to free a pebble. But it is still unnecessary
to check the neighbors of {\it both} ends --- this basically has to do with the fact
that any subnetwork containing two given sites and rigid by itself (thus
having only six associated free pebbles) always includes at least some neighbors
of {\it both} of these sites.

\begin{figure}
\begin{center}
\includegraphics[width=3in]{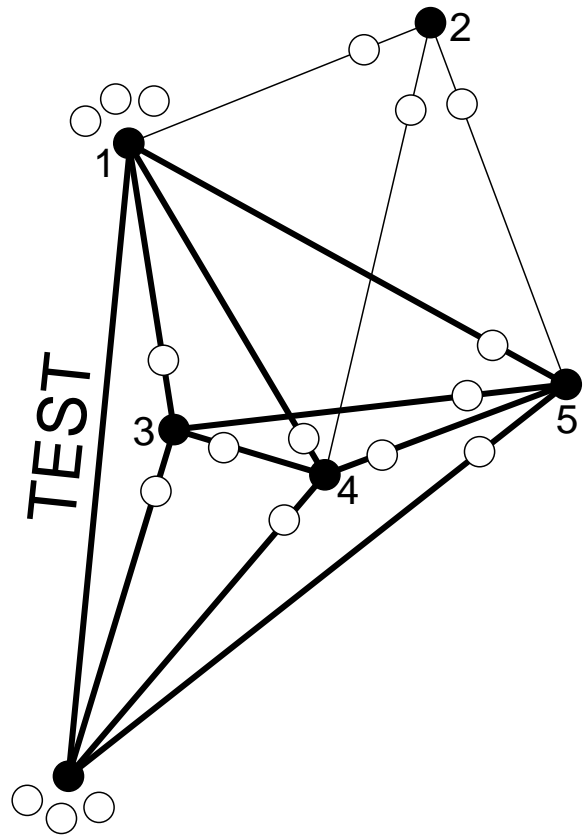}
\end{center}
\caption{An example of a network for which different failed pebble search regions
do not coincide and the intersection of the regions needs to be taken to
identify the stressed region correctly. The constraint marked ``TEST'' is being
inserted. Six pebbles are freed at its ends (shown). No other free pebbles are
present, so search for the seventh free pebble cannot succeed. An attempt is made
to free the seventh pebble at each neighbor of site 1. When this is done at the
site marked 2, sites 3, 4, 5 are all passed when searching for a pebble. But
when this is done at one of the sites marked 3, 4, 5, site 2 is not passed,
since the thin constraints leading to this site are not covered by pebbles belonging
to either of the sites 3, 4, 5. Clearly, only thick constraints
are stressed, and thus the stressed subgraph should not include site 2, so the
intersection of failed search regions needs to be taken to identify the stressed
region correctly.}
\label{intersec}
\end{figure}

There is an important difference
regarding the stressed region determination. It is no longer true that all
regions of failed pebble search for each of the neighbors of an end of the
constraint being tested are going to coincide. So even when failure to free an extra pebble
is detected for one of the neighbors, the procedure should still be repeated for
all of the other neighbors and the {\it intersection} (not the union!) of
the
regions of failed search is the new stressed region. Of course, we should remember that even this
more complicated procedure is not completely error-free: for instance, we still
assume that stressed regions are induced subgraphs, but, as explained in the previous
section, this is not necessarily true. The justification for the fact that the
intersection of the failed search regions should be taken is as follows.
Imagine a network consisting of just those constraints that are covered by a
pebble. As only independent constraints are covered, such a network will have
no redundancy and thus no stress. When a new constraint is inserted, a stressed
region appears if this constraint is redundant. Any constraint from such a
region can be removed without changing rigidity but making the stressed region
unstressed, while any constraint from outside the region will add one floppy
mode but the stress will remain.
So removing a constraint from inside the stressed region and freeing the
associated pebble should make this pebble available to every neighbor of the
ends of the newly inserted constraint (as this constraint should now become
independent) --- in which case the site to which the pebble belongs is part of
{\it all} pebble search regions; conversely, the pebble from any constraint from
outside the region should not become available to {\it at least one} of the neighbors
--- and then the site to which the pebble belongs is not part of at least one
of the pebble search regions. An example of a network where failed search
regions differ and clearly an intersection of those regions needs to be taken
is shown in Fig.~\ref{intersec}. Of
course, if implied hinges are present or would appear upon removal of a single
constraint, this procedure may not work correctly, as for the examples in
Figs.~\ref{2banana} and \ref{bananalocked}.

\begin{figure}
\begin{center}
\includegraphics[width=3in]{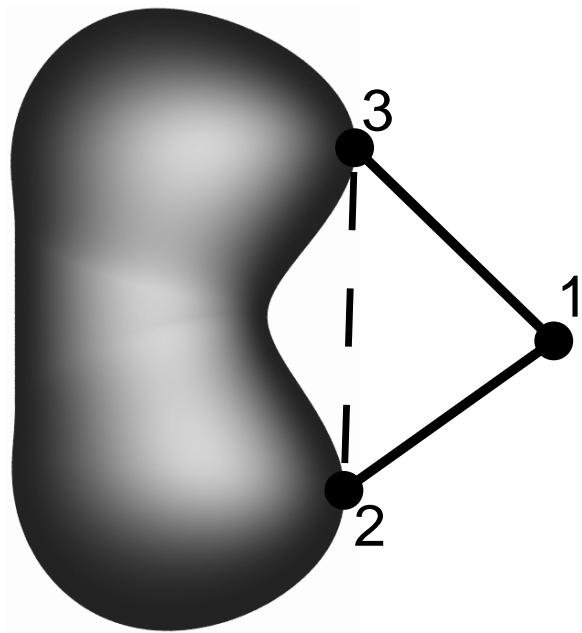}
\end{center}
\caption{A sketch of the simplest (``trivial'') implied hinge.
Constraints 1--2 and 1--3 are present, but constraint 2--3 is not. A body on the
left denotes a rigid cluster that is rigid by itself. The triple 1--2--3 is also
a rigid cluster, but is not rigid by itself. Line 2--3 is an implied hinge. In
many cases, this is the most frequent situation involving an implied hinge. It
will not cause errors in the floppy mode count or stress determination, but the
cluster 1--2--3 may be missed by the rigid cluster decomposition procedure.}
\label{trivial}
\end{figure}

Regarding rigid cluster decomposition, one serious issue is choosing a starting
set of three mutually rigid sites for each cluster. In the bond-bending case,
we started with a triple consisting of an arbitrary site (having at least two
neighbors) and two of its first neighbors knowing that they always form
a mutually
rigid set. Six free pebbles were then collected at these three sites. In the
general case, 
unfortunately, not every angle is rigid, i.e., not every triple
consisting of a site and its two neighbors is a mutually rigid set.
If we still do the same, i.e., choose an angle and collect as many pebbles as
possible at the three sites forming it, then sometimes it may be possible
to collect more
than 6 pebbles. It may be so because the set is not mutually rigid; but it
may also be so when the set is mutually rigid, but only because the rigid
cluster that it is a part of is rigidified from outside. So, if we rely on
the number of freed pebbles to determine if the angle is rigid, we may fail
to identify some of the rigid clusters. The simplest example is in
Fig.~\ref{trivial}. The
three explicitly marked sites in this figure, 1, 2 and 3, form a rigid cluster,
but since it is not rigid by itself, it will always be possible to collect 7
pebbles at these sites, and thus
this cluster will be missed. In this particular case, it is easy to detect the
error: if one inserts a constraint between sites 2 and 3,
this constraint is
redundant, and so sites 2 and 3 are mutually rigid and then all three sites are
mutually rigid. However, even if the failure is detected in such a way (and it
is not always possible), it is not obvious in general how to proceed from
there.
Do we keep all 7 pebbles free? Do we only free six of the pebbles?
Both of these choices are potentially problematic. So we
have chosen to limit ourselves
to the test based on the number of freed pebbles. That is, we choose an angle
and try to free as many pebbles as possible at the first site and then at both
of its neighbors keeping the previously freed pebbles free. If only 6 pebbles
are freed, the angle is deemed rigid and we proceed exactly as in the
bond-bending case. Otherwise, the angle is deemed non-rigid and another angle
is chosen instead. This means that we are going to miss rigid clusters such as
1-2-3 in Fig.~\ref{trivial}, which, however, is of minor importance for most purposes. Some other clusters can be missed, too,
but, at least in the
examples we consider later in this paper, this is much more rare. Note that
this problem can still only appear due to implied hinges (or
uncovered explicit hinges --- see Section~\ref{errors}), as we show when we
analyze possible errors of the pebble game in more detail in
Section~\ref{errors}. We should note that it is also possible to err on the other
side, i.e., to misidentify a non-rigid region as rigid, for instance, in the
double-banana case (Fig.~\ref{2banana}), where the whole network is deemed rigid by the
pebble game, but there is, in fact, a hinge.
Once a rigid angle is found, the
associated rigid cluster is mapped as in the bond-bending case. Then another
rigid angle is chosen, and so on, until all rigid angles are assigned to
clusters.

\section{The relaxation algorithm for exact rigidity analysis}
\label{relax}
In order to test the accuracy of the pebble game algorithm described in the
previous section, we need a way to do exact rigidity analysis. In this section,
we describe one possible method, which we call the {\it relaxation algorithm}.
Like more straightforward methods, such as numerical diagonalization of the
dynamical matrix or singular value decomposition (SVD) of the rigidity
matrix~\cite{svd}, our
approach is not an integer algorithm and thus potentially subject to round-off
errors (although the method incorporates several consistency checks which make
any errors in the final result unlikely). Unlike such straightforward
methods, the relaxation algorithm also relies on some facts from rigidity
theory, in particular, the Dress conjecture is used to find the number of floppy
modes. Also, unlike for the diagonalization and SVD procedures and similar to
the pebble game, the actual
eigenmodes (including the eigenvectors corresponding to the floppy motions) are
not obtained; on the other hand, rigid cluster decomposition is easier to
obtain using the relaxation algorithm. Based on this, it can be said that the
relaxation algorithm is ``intermediate'' between the straightforward
approaches and the pebble game.

Suppose we are given a network topology for which rigidity properties need to
be obtained. Consider a particular realization of that topology, i.e., an
elastic network (modeled as a network of harmonic springs) with specified
equilibrium positions of sites and
whose connectivity is consistent with the given topology. For the first stage
in the relaxation algorithm, used to obtain the rigid cluster decomposition and
the number of floppy modes, assume that the natural lengths of the springs
are chosen to fit exactly between the sites at specified positions, so that
initially the network is in equilibrium and unstrained, thus being at the
energy minimum. Now, displace all sites by infinitesimal amounts in
random directions. In general, the network will no longer be in equilibrium.
If the network is now relaxed using, for instance, the conjugate gradient
algorithm~\cite{congrad}, then after the relaxation is complete, the network is again in
equilibrium. However, generally speaking, the positions of sites will not
coincide with the initial equilibrium positions. This is because the equilibrium
is not unique: any displacement from the initial equilibrium corresponding to a
floppy motion will leave the energy unchanged and thus still equal to its
minimum value of zero. Thus we can expect the final configuration (after relaxation)
to differ from the initial one (before displacing the sites) by a
$3N$-dimensional vector that belongs to the space of floppy motions. Since
floppy motions do not change distances between mutually rigid sites, then
for any pair of mutually rigid sites, the final distances will be the same as the initial
distances. On the other hand, since the displacement was chosen at random,
it is very unlikely that for a pair of sites that are {\it not} mutually rigid,
the distances will be the same at the beginning and at the end. Thus the
procedure described above allows finding of all mutually rigid pairs of sites.
All displacements have to be sufficiently small (ideally, infinitesimal):
otherwise, the system can jump from one local minimum to another. If initial
displacements are infinitesimal, they will remain infinitesimal during and
after relaxation.

If the initial position of site $i$ is ${\bf r}_i$ and the infinitesimal
displacement from that position is ${\bf u}_i$, then to the lowest order in
$\{{\bf u}_i\}$, the change in the distance between sites $i$ and $j$ is
\begin{equation}
\delta r_{ij}=\frac{({\bf r}_j-{\bf r}_i)\cdot ({\bf u}_j-{\bf u}_i)}
{|{\bf r}_j-{\bf r}_i|}.
\end{equation}
Spring constants can be chosen arbitrarily, as the final result does not
depend on them. It is convenient to choose them so that the spring constant for
the spring between sites $i$ and $j$ is equal to $|{\bf r}_j-{\bf r}_i|^2$. Then
the total energy is
\begin{equation}
U={1\over 2}\sum_{\langle ij\rangle} [({\bf r}_j-{\bf r}_i)\cdot
({\bf u}_j-{\bf u}_i)]^2,
\end{equation}
where the sum runs over all pairs of sites that have a constraint between
them. After relaxation, for each pair of sites, $\{k,\ l\}$, the quantity
\begin{equation}
\delta_{kl}=({\bf r}_l-{\bf r}_k)\cdot ({\bf u}_l-{\bf u}_k)\label{scal}
\end{equation}
can be used to determine if the distance between these sites has changed:
if this quantity is zero, then the distance has not changed and these sites
are mutually rigid; otherwise the distance has changed and the sites are not
mutually rigid. Note that since $U$ is quadratic and $\delta_{kl}$ linear
in displacements $\{{\bf u}_i\}$, these displacements no longer have to be
infinitesimal when using these ``linearized'' equations: indeed, rescaling
all displacements by some arbitrary factor does not change the results. This
has an advantage that in the actual implementation of this procedure on a
computer, one does not have to worry if the displacements are ``small
enough'' (which would be the case if linearization was not done).

\begin{figure}
\begin{center}
\includegraphics[width=3in]{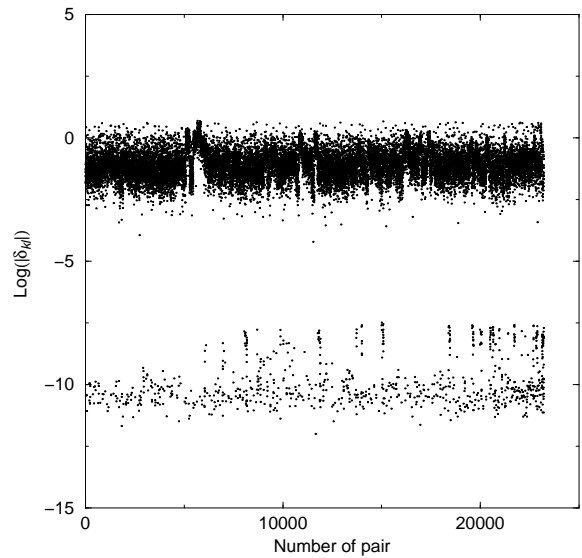}
\end{center}
\caption{For the relaxation procedure described in the text, the base-10
logarithm of the sum over two realizations of the absolute value
of the quantity $\delta_{kl}$ in Eq.~(\ref{scal})
for all pairs of sites of a network with
216 sites. The gap between ``zero'' and ``non-zero'' values is clearly seen.
Pairs with the values below the gap are mutually rigid, those with the values
above the gap are not.}
\label{scalar}
\end{figure}

In practice, the computer precision is always limited, of course, so in an
actual implementation, the values of $\delta_{kl}$ for rigid pairs found
numerically will be very small but non-zero. A possible solution is
introducing a {\it cutoff}: pairs of sites for which $\delta_{kl}$ is below the
cutoff are deemed rigid, and those for which $\delta_{kl}$ is above the cutoff
are not rigid. However, another complication is possible: if the
realization is accidentally very close to a non-generic one or a special initial
displacement was chosen, some of the non-rigid pairs may have the
corresponding value of $\delta_{kl}$ below the cutoff. For this reason, at
least two realizations are always run; the absolute values of $\delta_{kl}$ are
summed up and then it is determined if a gap of at least two decades containing
no values of $\delta_{kl}$ has formed. If it is, the cutoff is chosen inside the
gap and the procedure is finished. This situation is illustrated in
Fig.~\ref{scalar}. If the gap has not formed, another realization is run and the
absolute values of $\delta_{kl}$ are added to the previously obtained sums. If
the gap is now present, then the procedure is finished, if not, the current run
is abandoned, and a new run is started. A new run is also started if in any
of the relaxations, a certain low energy is not achieved in a predetermined
number of steps. At most three runs are done; if none are successful, the
relaxation procedure has failed; it can then be repeated using a higher
precision. Even despite always doing at least two runs and even when
the self-consistency checks described below succeed, there is still a very
small chance of an error. In cases of doubt, for instance, when a
discrepancy with the pebble game is detected, the procedure can be done
as many times as desired, and lower final energy tolerances and larger
gap sizes can be set.

Once all pairs of mutually rigid sites are found, the next step is finding all
hinges, including implied ones. First, make a list of sites such that among
their neighbors, not all are mutually rigid --- only such sites can be hinge
endpoints. From this list, choose all pairs of mutually rigid sites.
For each such pair, \{A,B\}, select an arbitrary site C rigid with respect to
both A and B. Then go through the list of all other sites rigid with respect
to both A and B; if {\it any} of such sites is {\it not} rigid with respect to
C, then A--B is a hinge (an implied one if there is no constraint A--B in the
network).

If all implied hinges are added to the
network explicitly as constraints, the configuration of rigid clusters in the
network and the number of floppy modes are not affected. But, as follows from
the Dress conjecture, all rigid clusters become contiguous. So once all
implied hinges are identified, it is convenient to add them to the network as
constraints and then mark rigid clusters labeling each rigid angle formed by
network constraints (including the just added implied hinges) so that angles
belonging to the same cluster are labeled
identically. Recall that sites forming an angle (or any triple of sites for that
matter) can belong to at most one rigid cluster, so that each rigid angle will
be assigned just one label. Because of contiguity of rigid clusters, such
labeling retains the full rigidity information; it is always possible to
traverse the network between any two mutually rigid sites by going through
angles assigned the same label. This can be used as a self-consistency check
for the algorithm. Constraints that do not form any labeled (i.e., rigid) angles are not
rigid with respect to any other sites and form single-constraint clusters on their
own; likewise, isolated (disconnected) sites are considered single-site
clusters. While finding all mutually rigid pairs of sites does not involve any
assumptions from the rigidity theory, the decomposition into rigid clusters,
as well as the self-consistency check using this decomposition, do rely on the
Dress conjecture.

The next stage is finding the number of floppy modes. The procedure,
as described below, is just a convenient interpretation of the Dress conjecture
that gives the number of floppy modes based on the number of (explicit and
implied) hinges~\cite{tibor}. Intuitively, it follows from the assumption that
once implied hinges are added, all clusters become rigid by themselves and
redundant constraints in each of them can be counted separately and then added
up. Any cluster of 3 sites
or more should have 6 floppy modes when isolated from the rest of the
network; then, if it contains $n$ sites and $c$ constraints, the number of
redundant constraints [according to Eq.~(\ref{Maxexact}) with $N=n$, $N_c=c$ and
$F=6$] is $6-3n+c$. Obviously, the numbers of redundant constraints should be
non-negative for all clusters and this serves as another self-consistency check
for the algorithm. After the numbers of redundant constraints are found for
each cluster, these numbers are added up. Note that even
though some constraints (namely, hinges) belong to two or more clusters
simultaneously, they are included in the count of constraints for
each cluster they belong to when obtaining the numbers of redundant constraints
for these clusters. The total number of redundant constraints $N_R$ is
then used in Eq.~(\ref{Maxexact}) to obtain the number of floppy modes $F$. 

Counting redundant constraints in each rigid cluster gives information on the 
presence or absence of stress within that cluster. However, even if the presence of
stress is detected, the information obtained so far does not indicate where
exactly it is located within the cluster. Remember that we have specifically
chosen the lengths of constraints so that they fit exactly, so there is no
stress in our networks after relaxation even when it has to be present
generically. To locate stress, we need to carry out another relaxation
procedure, this time with constraints that do not fit exactly. In the spirit
of the first relaxation procedure, we choose constraints with
{\it infinitesimal} misfits. That is, just as in the first procedure, sites are
first assigned random positions in space, ${\bf r}_i$; if, according to the
given connectivity table, there is a constraint between sites $i$ and $j$, its
length is chosen equal to $l^0_{ij}=|{\bf r}_j-{\bf r}_i|+\Delta_{ij}$, i.e.,
there is an infinitesimal misfit $\Delta_{ij}$. As the misfits are
infinitesimal, it is expected that displacements from the initial position,
${\bf u}_i$, will also remain infinitesimal at all times during the subsequent
relaxation procedure, as well as in the relaxed network. In the lowest order in
${\bf u}_i$, the deformation of the constraint between sites $i$ and $j$ is
\begin{equation}
\delta l_{ij}=\frac{({\bf r}_j-{\bf r}_i)\cdot ({\bf u}_j-{\bf u}_i)}
{|{\bf r}_j-{\bf r}_i|}-\Delta_{ij}.
\end{equation}
If we choose the spring constant equal to $|{\bf r}_j-{\bf r}_i|^2$ and
introduce $\epsilon_{ij}=\Delta_{ij}|{\bf r}_j-{\bf r}_i|$, the energy is
\begin{equation}
U={1\over 2}\sum_{\langle i,j\rangle}
[({\bf r}_j-{\bf r}_i)\cdot ({\bf u}_j-{\bf u}_i)-\epsilon_{ij}]^2.
\label{strpot}
\end{equation}
As in the first stage of the relaxation algorithm, we displace all sites at
random initially, although now this is not really important, since constraint
lengths are themselves random and do not fit exactly between sites.
After relaxation with the potential (\ref{strpot}) is done, quantities
\begin{equation}
\delta_{ij}=({\bf r}_j-{\bf r}_i)\cdot ({\bf u}_j-{\bf u}_i)-\epsilon_{ij}
\label{str}
\end{equation}
are used to find stressed constraints: if $\delta_{ij}=0$, then the constraint
between sites $i$ and $j$ is unstressed, otherwise it is stressed. In practice,
a cutoff between ``zero'' and ``non-zero'' values is again introduced. 
Similarly to the first relaxation procedure, since $U$ is quadratic and
$\delta_{ij}$ linear in both $\epsilon_{ij}$ and ${\bf u}_i$, 
these two latter quantities
do not have to be infinitesimal when using these ``linearized'' equations.

Finally, we analyze the computational speed of the relaxation algorithm. In
theory, the conjugate gradient algorithm converges to the exact minimum after
the number of steps equal to the number of degrees of freedom, which is $3N$
for a network of $N$ sites, or ${\cal O}(N)$. Each of these steps requires the
evaluation of the gradient of the potential, which takes ${\cal O}(N)$
floating-point operations (flops), so that the relaxation procedure proper
takes ${\cal O}(N^2)$ flops overall. In practice, away from the rigidity percolation
transition the number of steps required may be significantly
smaller; on the other hand, very close to the
transition a somewhat larger
number of steps may be needed, since because of the round-off errors the
convergence is not perfect after $3N$ steps. Another potentially costly
part of the algorithm is finding hinges. Formally, this part requires
${\cal O}(N^3)$ operations; but in practice, for typical network sizes (up to
a few thousand sites), it is usually much faster than the relaxation proper,
which in part is explained by the fact that only integer operations are
involved. In fact, it may be possible to optimize this part so that in
practice, it is never (or rarely) ${\cal O}(N^3)$.

These estimates of the computational cost of the relaxation algorithm should be
compared to those for diagonalization and SVD. Since the computational cost of
both diagonalization and SVD is ${\cal O}(N^3)$ (if the number of constraints
is comparable to the number of degrees of freedom)~\cite{golub}, it is likely
in view of the above analysis that the relaxation algorithm is faster (although
direct tests need to be done to find out if this is really so for typical
problems and network sizes). The advantage of
the more traditional approaches is, as mentioned before, the fact that they
also find the floppy modes themselves, not just their number; on the other hand,
rigid cluster decomposition is problematic. In fact, even if a traditional
approach is used, we suggest combining it with a procedure similar to ours to
find rigid clusters. Namely, once floppy modes are obtained, their linear
combination with random coefficients can be used as the analog of the outcome
of relaxation; then the values of $\delta_{kl}$ can be calculated using
Eq.~(\ref{scal}) and the rest
of the procedure for finding rigid clusters is the same as described above.
Note also that in the case of diagonalization and SVD, a cutoff still needs to
be defined between ``zero'' and ``non-zero'' eigenvalues or singular values;
but note that there are only ${\cal O}(N)$ of these values, whereas there are
${\cal O}(N^2)$ values of $\delta_{kl}$ used to define the cutoff in the
rigidity part of the relaxation algorithm, and the more values are used to
define the cutoff, the more confident can one be that the cutoff is chosen
correctly. This is another advantage of the relaxation algorithm, although,
unfortunately, it does not apply to the stress determination part, since in this
case, the number of values of $\delta_{ij}$ from Eq.~(\ref{str}) is equal
to the number of constraints and thus is ${\cal O}(N)$.

To conclude this section, we should mention certain similarity between the
relaxation algorithm for rigid cluster decomposition and the FRODA algorithm for
generating internal motions of proteins~\cite{froda, site}. In both algorithms,
the initial configuration is distorted in some way and then relaxed to generate
a new configuration. The details are, of course, different: in the relaxation
algorithm, the displacements are effectively infinitesimal, in FRODA, they are
not; FRODA uses a different relaxation procedure that makes use of special
properties of protein networks; also, as a consequence of dealing with finite
displacements, FRODA needs to take care of finite atom sizes by avoiding van der
Waals overlaps.

\section{Errors in the pebble game: general considerations}
\label{errors}
We are now in a position to analyze the correctness of the approximate pebble
game described in Section~\ref{peb} in different situations. We start with
some general considerations.

First of all, consider what configurations can give rise to pebble game errors.
As mentioned before, according to the Dress conjecture, all violations of the
molecular framework conjecture statement and the associated properties of
contiguity of rigid clusters and their being rigid by themselves are
due to implied hinges. However, it is important to remember that during the
pebble game, the network is built gradually, constraint after constraint, and
the counting of redundant constraints is done during the whole construction
process. So even if the final network does not contain implied hinges, but they
were present at some time during the construction process, there still may be
errors.

Consider first the situation where an implied hinge becomes explicit when a
constraint coinciding with it is inserted. This constraint is always
redundant and normally will not be covered by a pebble during the pebble game.
(Exceptionally, the pebble game may deem it independent incorrectly and cover it because
of errors due to hinges elsewhere in the clusters sharing the given hinge, but
in this case the end results of the pebble game are the same as if the
constraint was inserted early enough so there was never an implied hinge in its
place, so we do not consider this
situation separately.) But any constraint not covered by
a pebble is ``invisible'' to the pebble game in the sense that further pebble
searches will occur in the same way and with the same outcome as if this
constraint was not present in the network. For this reason, an explicit hinge
coinciding with an uncovered constraint will, in fact, always cause the same
problems as if it was an implied hinge. Note that this is so even if the hinge
was {\it never} an implied one.

Another potentially troublesome situation would be a hinge
that existed at some time during the construction process but ceased to exist
altogether when the rotation around it was locked later. In this case, however,
if there was an error in the floppy mode counting due to the hinge, it would
be corrected when the hinge is locked. For instance, the floppy mode count
given by the pebble game for the double-banana graph in Fig.~\ref{2banana} is 6, one less
than the actual number. However, 6 is the lowest possible floppy mode count, as
it corresponds to the rigid body, so any additional constraints, including those
locking the rotation around the implied hinge (such as the thin constraint
in Fig.~\ref{bananalocked}) will be (incorrectly) deemed
redundant, the two errors in the counting will cancel out and the counting will
stay at 6, which is the correct answer when the
hinge is locked. As for rigid cluster decomposition, it is done at the end of
the pebble game; so if the locking constraint is covered (as would be the case,
e.g., when the hinge in Fig.~\ref{2banana1} is locked), there are certainly no
problems, (as there is no difference in pebble arrangement compared to the
case when the locked constraint is inserted even before the hinge had a chance
to appear);
if, on the other hand, it is deemed redundant and not covered, this means that
the two clusters whose mutual rotation the constraint is supposed to lock
were (incorrectly) found mutually rigid even before locking, so again
in the end there are no problems. So the conclusion is
that there are no problems in either the floppy mode count or the rigid cluster
decomposition due to hinges that were there but are not there anymore at the
end. But the stress determination may still be wrong in such cases, as
Fig.~\ref{bananalocked} illustrates.

When an implied or an uncovered explicit hinge is present, it may affect both
the floppy mode count and the rigid cluster decomposition (as in
Fig.~\ref{2banana}), or it may only affect the latter (as in
Fig.~\ref{2banana1}). To find out what determines the difference, recall that
according to the Dress conjecture, when obtaining the number of floppy modes
$F$, redundant constraints are counted in each cluster separately (after all
implied hinges are added) and then summed up. When counting is done within a
given cluster, the hinge contributes to the number of redundant constraints
(that is, its addition changes this number) if it is stressed
when this cluster (with the hinge included explicitly) is taken separately from
the rest of the network (with a caveat mentioned at the end of the paragraph).
So the contribution of a hinge to the total number of
redundant constraints is equal to the number of times this hinge is found stressed
when different clusters sharing this hinge are taken in isolation. On the other
hand, the hinge always contributes 1 to the total number of constraints. So
the net contribution of the hinge to $F$ is the number of times the hinge is
found stressed minus one. Any implied or uncovered explicit hinge has to be
found stressed at least once. If it is found stressed once, the net contribution
is zero. If it is found stressed more than once, the net contribution is
non-zero. In the pebble game, on the other hand, implied hinges do not
contribute to either the total number of constraints or the number of redundant
constraints, so the net contribution is always zero; as mentioned above,
uncovered explicit hinges are equivalent to implied ones for floppy mode count
purposes, so the net contribution is zero for them as well. The conclusion is
then that when an implied or uncovered explicit hinge is found stressed once,
the floppy mode count of the pebble game is correct. This is the case in
Fig.~\ref{2banana1}, where
the hinge is found stressed when the left banana is considered, but not when the
right banana is considered. On the other hand, when a hinge is found stressed
more than once, the pebble game count is incorrect (always lower than
the correct one, never higher). In Fig.~\ref{2banana}, indeed, the hinge is
found stressed twice, i.e., it is found  stressed when considered with each of
the two bananas. Occasionally, the considerations of this paragraph may
{\it overestimate} the number of redundant constraints and thus the pebble game
error due to implied hinges, if there are several hinges within the same
rigid cluster and the same stressed region. For instance, if there are two
hinges within the same stressed region, then according to the above
considerations, the contribution of these hinges to the number of redundant
constraints in the cluster will be equal to 2; but it is still possible that
the actual constraint count for this region will indicate that it has just one
redundant constraint.

Note that since the relaxation algorithm finds explicit and implied hinges and
also counts redundant constraints within each cluster, we can obtain some
information on the possible error in the pebble game floppy mode count even
without running the pebble game. Consider each hinge (explicit or implied) and
determine the number of clusters among those that share this hinge that have a
non-zero number of redundant constraints $n_r$. Having a non-zero $n_r$
is a {\it necessary} condition for the hinge to be found stressed when
considered with this cluster. But it is not a sufficient condition, since a
non-zero number of redundant constraints only indicates that stress is present
somewhere in the cluster, and this does not necessarily include the hinge. For
this reason, if the number of clusters sharing the hinge and having non-zero $n_r$
is zero, the hinge is unstressed; it is then explicit and moreover, is covered,
so it can never spoil finding the number of floppy modes or the configuration of rigid
clusters. If the number of clusters with non-zero $n_r$ is 1, the actual number of
times the hinge is found stressed is either 0 or 1, and then the hinge is either
not dangerous at all (in the first case), or can affect the rigid cluster
decomposition (in the second case, provided that it is implied or inserted late
enough to be uncovered); it cannot affect the floppy mode count. Finally, if the
number of clusters with non-zero $n_r$
is 2 or higher, it is possible that the hinge is stressed two or more times, and
then the number of floppy modes {\it may} be affected. The maximum possible
error in the number of floppy modes due to the hinge is the number of clusters
with non-zero $n_r$ minus 1. The advantage of the described procedure of finding
dangerous hinges (as opposed to the straightforward comparison of the pebble
game results with the relaxation results) is that the error determined in this
way is at least as high as the maximum possible error (where the maximum is
taken over all possible orders of constraint insertion). In other words, this
is the worst-case scenario estimate.

Regarding stress determination, as mentioned above, there may be problems if a
single constraint locks a mutual rotation of two clusters around a hinge, as in
Fig.~\ref{bananalocked}. Unfortunately, since in such cases the hinge is not actually present,
it is impossible to find such situations, except by doing stress determination
through relaxation and then comparing directly with the pebble game result.
Another case when problems with finding stress can arise is when the floppy mode count fails, like
in the double-banana case. We should note that problems with stress
determination can only arise due to hinges or ``former'' (i.e., locked) hinges
that can create problems for the floppy mode count, but not those that can
only affect rigid cluster decomposition. This is because the stress
determination procedure is designed to be correct if there are no errors in the
floppy mode count for either the given network itself or any network obtained
from it by removing a single constraint.

\section{Randomly diluted central-force networks}
\label{CFnets}
In this section, we consider a particular class of 3D networks, randomly
bond-diluted central-force networks, using the algorithms considered in
the previous sections. In particular, we study rigidity percolation on
such networks. We first apply the
relaxation algorithm to the networks in order to ascertain the possibility of
their study with the faster pebble game algorithm, whose use allows studying
much larger networks than would be feasible with the relaxation algorithm. At
the end of the section, we briefly consider {\it site-diluted} networks, with
some of the results markedly different from those for
bond-diluted networks.

\subsection{Maxwell counting}
We start with the straightforward Maxwell counting in order to guide our search
for the rigidity transition.

In a network of $N$ sites, the number of degrees of freedom is $3N$.
In central-force networks, the number of constraints is equal to the number of bonds and
is $\ra N/2$, where $\ra$ is the mean coordination (the average number of
sites connected to a given site). Then
\begin{equation}
F_{\rm Maxw}=3N-\ra N/2.
\end{equation}
This reaches zero at $\ra = 6$. Even though Maxwell counting is not exact, it is
assumed that in reality, the number of floppy modes becomes small when $\ra$
approaches this value and a percolating rigid cluster emerges somewhere around
this point (indeed, in 2D central-force and 3D
bond-bending networks, the transition occurs very
close to the point at which the Maxwell counting result turns zero).

Given that the transition is expected to be located at $\ra\approx 6$, to study
it by bond-diluting a regular lattice we need
a lattice with the coordination number exceeding 6. Thus the body-centered cubic
(BCC) lattice with the coordination of $z=8$ and the face-centered cubic (FCC)
lattice with the coordination of $z=12$ are natural choices.
In a regular lattice with coordination number $z$, the total number of bonds is
$Nz/2$. If bond dilution is done so that fraction $p$ of the bonds remain,
this gives the number of bonds $N_B=Nzp/2$, and since each bond is shared
between the two sites that it connects, the mean coordination is
\begin{equation}
\ra=2N_B/N=zp.\label{rp}
\end{equation}
Then at the transition we expect $p\approx 1/2$ for FCC and $p\approx 3/4$ for
BCC. Note, by the way, that Eq.~(\ref{rp}) is valid even for finite networks and
even non-random ones, if $p$ is indeed interpreted as the actual fraction of
present bonds and not as the probability that a given bond is present.

One should keep in mind that the approaches to studying rigidity described in
this paper are only applicable to {\it generic} networks. Regular lattices like
FCC and BCC are not generic, of course, as they have parallel bonds, all bonds
are of the same length, etc. Any results described here are therefore applicable
not to the diluted regular lattices themselves, but rather to networks
topologically equivalent to them, but distorted by introducing bond length
disorder. This is also true for the older results for the diluted diamond lattice
bond-bending networks shown in Fig.~\ref{perc3DBB}.

\subsection{Using the relaxation algorithm}
\label{relaxCF}
We first study small bond-diluted BCC and FCC networks using the relaxation algorithm, doing
both rigidity and stress runs (the latter just for those networks
where the former detected any redundancy). For
FCC, we use networks of 500 sites, for all numbers of bonds
between 1460 ($\ra=5.84$ or $p\approx 0.4867$) and 1490 ($\ra=5.96$ or
$p\approx 0.4967$). For each number of
bonds, we generate 100 different networks; this gives a total of 3100 networks.
Even though usually much larger networks can be analyzed easily, this is a
particularly difficult case for the relaxation algorithm, since close to the
transition, there are large regions (taking up most of the network) that
are isostatic (i.e., rigid but unstressed, with constraints exactly balancing
degrees of freedom) or nearly isostatic, and it is very hard computationally
to distinguish an exactly isostatic region from one lacking just a single
constraint and thus having one floppy mode spread over thousands of degrees of
freedom: in the former case, the region is rigid but with extremely low
effective elastic moduli (vanishing in the thermodynamic limit); in the latter
case, it is flexible but with the motion limited to a subspace of dimensionality
one in a space with thousands of dimensions. For this reason, in rare cases the
relaxation algorithm has failed to converge. Namely, in the rigidity runs, out of
3100 networks, relaxation has failed for 39, or just over 1\%. All these cases
were re-run using more computationally intensive quadruple-precision arithmetic;
all 39 runs succeeded. Likewise, in the stress runs, relaxation has failed in
23 cases, and again, all succeeded using quadruple precision. At the lowest
bond number (1460), all 100 networks have only small rigid clusters and no
stress; at the highest bond number (1490), all 100 networks have a rigid cluster
taking up almost all the network and 99 out of 100 networks have a stressed
region likewise taking up most of the network. In other words, the chosen range
of mean coordinations indeed contains the rigidity transition. In the BCC case,
networks of 686 sites were used, with numbers of bonds between 2040
($\ra\approx 5.9475$ or $p\approx 0.7434$) and 2060 ($\ra\approx 6.006$ or
$p\approx 0.7507$), again
with 100 networks for each number of bonds, for the total of 2100 networks. In
this case, all runs, both for rigidity and for stress, succeeded without using
quadruple-precision arithmetic. Again, at the smallest bond number, only small
rigid clusters are present and no stress in all networks; at the highest bond
number, all 100 networks have a rigid cluster and a stressed region taking up
most of the network.

A remarkable observation is that in all these runs, for both FCC and BCC, only
very small and very large rigid clusters are observed, but
never those of intermediate size. Namely, in FCC networks, only clusters of
11 and fewer sites or 431 and more sites are found; in BCC networks, only
clusters of 1, 2, 3 sites or at least 668 sites are found. As for stressed
regions, small ones are never observed: the smallest regions ever found are of
size 388 for FCC and 632 for BCC. This is a strong indication that the
rigidity transition is {\it first order}: instead of the average cluster size
growing gradually as the transition is approached, before a
percolating cluster (that first takes up a small part of the network) arises,
here no gradual growth is observed; the percolating cluster emerges suddenly,
upon a single bond addition, and immediately takes up much of the network.
Likewise, stress is not present at all in the floppy phase and arises suddenly
after a single bond is added, again, taking up most of the network.

As we have mentioned, a first order transition was found previously in
so-called random bond networks (RBNs). In RBNs, sites
are assigned certain coordination numbers and then are connected at random,
regardless of the distance between them, consistent with the assigned
coordinations. The reason why the transition is first order in this case has to
do with the
absence of finite rings and thus finite rigid clusters in RBNs (other than
single bonds and single sites with their associated constraints in the
bond-bending case).
Without finite clusters, the infinite cluster
has to emerge suddenly: there is no diverging correlation length typical of
second order transitions as the threshold is approached. Of course, rings are
certainly present in regular lattices like BCC and FCC, but still, is the situation
similar here in some way? In the Appendix, we analyze this question in detail using a
computational procedure for generating maximally rigid configurations of a
certain size. The result is that in the BCC case, the situation is indeed
somewhat similar: besides the clusters of sizes 1, 2 and 3, the smallest
possible cluster has size 90 if standalone or 84 if sharing a hinge with another
rigid cluster (and these are extremely rare; the frequency per site of
observing such a cluster is estimated in the Appendix to be very roughly
$\sim 10^{-30}$);
but in the FCC case, clusters of {\it all} sizes are possible,
but the probability of actually observing a cluster of a given size decreases
very rapidly as the size increases.

We can now address the question of the accuracy of the pebble game algorithm
when applied to bond-diluted central-force BCC and FCC networks. In the BCC case,
when besides the percolating cluster, only clusters of size up to 3 are
observed, obviously the only possible kind of hinge is a ``trivial'' one
shared between the percolating cluster and the one of size 3 (the ``triangle''),
as shown in Fig.~\ref{trivial}. Indeed, such hinges are rather frequent: a
total of 4822 are observed in the 1276 percolating networks (out of the total of
2100 networks), including two that are shared with the percolating cluster by
{\it two} ``triangles''. All of these hinges are implied: indeed, triangles of
bonds do not exist in BCC lattices, so rigid ``triangles'' actually consist of two
``real'' bonds and an implied hinge. Such trivial hinges are also by far the
most frequent kind in the FCC case, although in this case, both implied and
explicit ones are possible: there are 15179 implied and 3971 explicit
``trivial'' hinges in 1288 percolating networks, including respectively 11 and
5 shared by two triangles. Besides these, there are also a very small number
(13 implied and 1 explicit) of hinges shared by the percolating cluster with
a {\it four-site} cluster (a tetrahedron); also, the two largest observed
``small'' clusters (of
sizes 10 and 11) share a few hinges with triangles. While the total number of
hinges is large, none of them can affect the floppy mode count of the pebble
game. Such hinges also cannot affect determination of the size of the
percolating cluster: indeed, since neither the triangle nor the tetrahedron
are stressed objects, they cannot rigidify the percolating cluster and it has
to be rigid by itself (cf. Fig.~\ref{2banana1}, where the left banana, although
not stressed by itself, becomes stressed if the hinge is inserted explicitly and
thus has the potential to rigidify the right, ``incomplete'' banana), thus the
pebble game will have no problem finding the percolating cluster correctly. The
only possible type of error is then failure to identify a rigid triangle or,
very rarely, tetrahedron. Since so few types of errors are possible, these
errors can easily be taken care of, if needed; but if we are only interested in
the number of floppy modes and the percolating cluster size (the quantities
relevant for determining the order of the transition), we need not do this.
Note also that the above numbers for problematic explicit hinges are likely
overestimations, as the rigid cluster decomposition part of the relaxation
algorithm overestimates the number of stressed hinges, as explained above. The
actual number can be found if required.

Obviously, since clusters of all sizes are possible in the FCC case, other
situations involving larger ``small'' clusters are possible in principle. For
example, one can ask if configurations of the double-banana type that would
violate the floppy mode count of the pebble game can occur. A method described
in the Appendix allows generation of such configurations, even if they are
extremely rare in reality. It turns out that the ``standard'' double-banana
graph (like the one shown in Fig.~\ref{2banana}) that consists of two clusters
(``bananas'') of size 5 is not possible in an FCC network. However, larger
configurations of the same type are possible. The smallest one consists of
two ``bananas'' of size 8 each, as shown in Fig.~\ref{8plus8}. In the
Appendix, the frequency of a single 8-site rigid cluster is estimated. It is
rather low, only $3\times 10^{-5}$ per site. To have a double-banana-type configuration,
two such clusters need to be located next to each other and in a certain
relative orientation; the probability of this will be roughly
$(3\times 10^{-5})^2\sim 10^{-9}$ per site ---
a very small number, and even this is probably an overestimation. Another
possibility for a double-banana-type configuration is the percolating cluster
sharing a hinge with a small cluster, the smallest possibility being a
six-site cluster (an octahedron), with the hinge connecting two opposite
vertices. Note that even though we have observed small clusters of sizes up to
11, such larger clusters are only present in the floppy phase, when there is no
percolating cluster. The largest  clusters coexisting with the percolating
cluster that we have seen are of size 4. Clusters of size 6 have never been
seen, and it is clear that the probability of seeing such a cluster is much
higher than that of seeing one attached specifically at opposite vertices
to the percolating cluster. In the Appendix, we give a crude estimate for the
frequency of hinges of this kind --- at most
$10^{-9}$ per site and probably much less --- again, extremely rare.

\begin{figure}
\begin{center}
\includegraphics[width=3in]{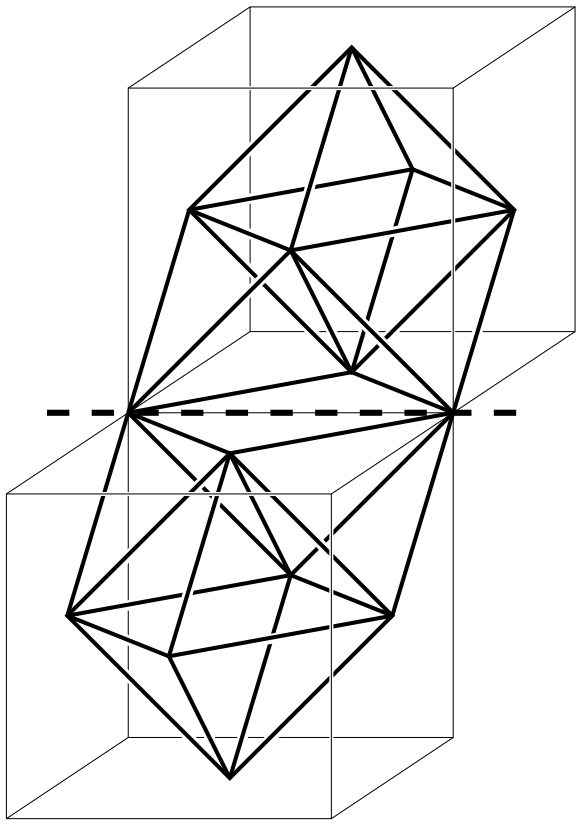}
\end{center}
\caption{The smallest graph of the double-banana type that can exist in the
diluted FCC network. The dashed line is the hinge. The outlines of two unit
cells are shown for clarity with thin lines.}
\label{8plus8}
\end{figure}

Regarding stress determination, as we have mentioned, finding hinges for
a particular network alone does not detect all possible errors in finding
stress using the pebble game because of configurations with locked hinges,
like the one in Fig.~\ref{bananalocked}. However, since we have never seen any
hinges that cause
problems for the floppy mode count in any of the networks we have analyzed, the
chance of having a locked hinge is very small --- there are no hinges to lock,
to start with. For this reason, we do not expect to see errors in finding
stress either.

\subsection{The pebble game analysis}

\begin{figure}[p]
\begin{center}
\includegraphics[width=3in]{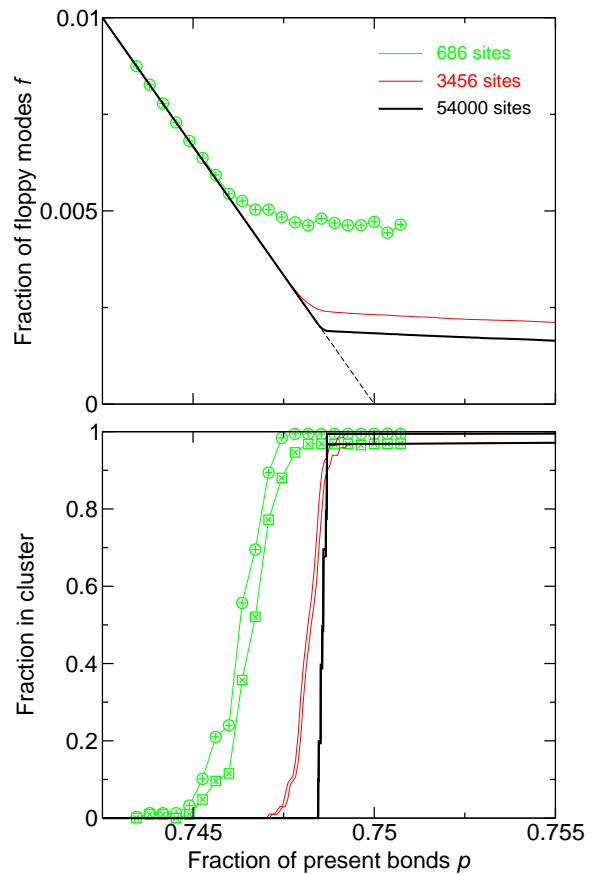}
\end{center}
\caption{(Top) The number of floppy modes per degree of freedom,
$f=F/3N$, for bond-diluted central-force BCC networks, for three different
sizes: small (686 sites; symbols, green online); medium (3456 sites;
the thin line without symbols, red online); and large (54000 sites;
the thick line). For the smallest size,
the results obtained by both the pebble game (circles) and the relaxation
algorithm (pluses inside the circles) are shown; the same realizations are used in
both cases. For the other two sizes, the pebble game was used. The
dashed line is the Maxwell counting result. (Bottom) The
fraction of sites in the largest rigid cluster and the fraction of
bonds in the only stressed region for bond-diluted central-force BCC networks,
for three different sizes. For each size, the top line represents the largest
rigid cluster and the bottom line, the stressed region. The line thicknesses
and the color scheme (in the online version) are the same as in the top panel.
For the smallest size, the results obtained by both the pebble game (circles
for the largest rigid cluster, squares for the stressed region) and the
relaxation algorithm (pluses and x's, respectively) are shown. For the
other two sizes, the pebble game was used. In both panels, the results
for the small and medium size are averages over 100 networks, with intervals between
data points equal to one bond; the results for the largest size are averages over
10 networks, with intervals between data points equal to one bond in the
vicinity of the transition and 10 bonds elsewhere.}
\label{BCC}
\end{figure}

\begin{figure}
\begin{center}
\includegraphics[width=3in]{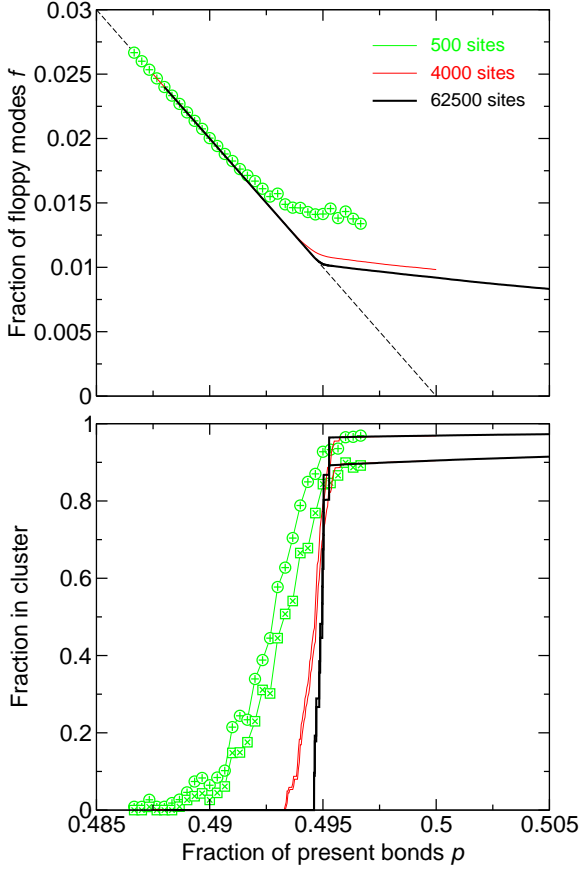}
\end{center}
\caption{Same as in Fig.~\ref{BCC}, for bond-diluted FCC networks. The sizes
used are 500 sites (small), 4000 sites (medium), and 62500 sites (large).}
\label{FCC}
\end{figure}

With the relaxation algorithm, we can only study rather small networks. While
this study gives strong indications that the rigidity transition in 3D
central-force bond-diluted networks is first order, using larger networks is
desirable, in particular, to reduce finite-size effects. Based on our results
obtained using the relaxation algorithm, we can be confident that the pebble
game results are going to be accurate in this case, even though in general the
algorithm is only approximate. We have also confirmed this explicitly, by
applying the pebble game to the same networks that we have analyzed using the
relaxation algorithm, as described in the previous subsection. The results of
this comparison (among other results described below) are shown in
Fig.~\ref{BCC} (for the BCC lattice) and in Fig.~\ref{FCC} (for the FCC
lattice). In the upper panels, we plot the average number of floppy modes
obtained using the relaxation algorithm (pluses) and the pebble game (circles).
In the lower panels, we show the average sizes of the largest rigid cluster
(pluses for the relaxation algorithm, circles for the pebble game) and of the
only stressed region (x's for the relaxation algorithm, squares for the
pebble game). All pluses are inside circles and all x's are inside squares,
indicating that the pebble game and the relaxation results coincide. In fact,
the results were checked network by network; very minor and rare discrepancies
(2 networks out of 3100 for FCC, none for BCC) turned out to be due to round-off
errors in the relaxation algorithm, rather than any problems with the pebble
game (as we confirmed by using quadruple precision on these networks that
eliminated the discrepancies).

Having ascertained the accuracy of the pebble game, we can now apply it to study
larger networks.
For the relaxation study, networks were generated independently at
each mean coordination. Here instead, networks are built gradually and
intermediate stages are used for obtaining results as well. We first remove all
bonds from the full FCC or BCC lattice and then place them back one by one
randomly while testing each for
redundancy with the pebble game. Rigid cluster decomposition is done after
every bond addition close to the transition, but can be done less frequently away
from it. In this way, we can analyze the whole
sequence of networks with different $p$ in a single pebble game run, which is
yet another advantage of the pebble game compared to other algorithms.

The results for the BCC and FCC lattices are shown in Figs.~\ref{BCC}
and \ref{FCC}, respectively. Lattice sizes used are 3456 and 54000 sites for BCC and
4000 and 62500 sites for FCC. In both cases, the percolation transition is clearly seen as a jump
in the size of the percolating rigid cluster and percolating stressed region, as
well as a break in slope in the number of floppy modes (which coincides with
the Maxwell counting result below the transition, as there are no redundant
constraints, but deviates immediately above the transition). The jump gets
sharper as the network size increases. This is consistent
with the rigidity transition being {\it first order}. The transition occurs at
$p\approx 0.7485$ in the BCC case and at $p\approx 0.495$ in the FCC case.
Both values are quite close to the Maxwell counting estimates ($p=3/4$ and
$p=1/2$, respectively) as expected. Note that there is a slight discrepancy
between our value in the BCC case and that obtained by Arbabi and
Sahimi~\cite{sahimi} ($p=0.737\pm 0.002$), which is probably due to the fact
that their simulations were on undistorted non-generic lattices.

\begin{figure}
\begin{center}
\includegraphics[width=3in]{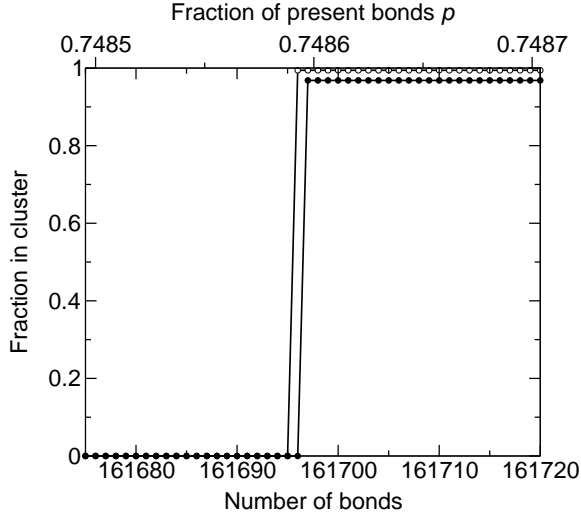}
\end{center}
\caption{The fraction of sites in the percolating rigid cluster (open circles)
and the fraction of bonds in the only stressed region (filled circles) as a
function of the number of bonds in the network, for a single bond-diluted BCC
network of 54000 sites. Note that the interval between adjacent data points is
one bond.}
\label{BCC_bond_one}
\end{figure}

\begin{figure}
\begin{center}
\includegraphics[width=3in]{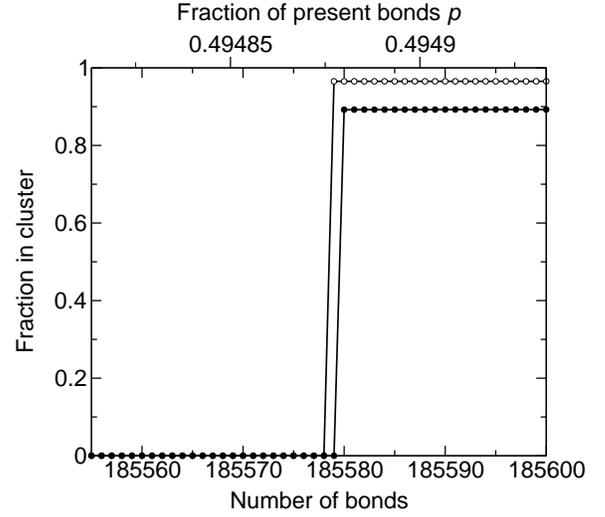}
\end{center}
\caption{Same as in Fig.~\ref{BCC_bond_one}, for a single bond-diluted FCC
network of 62500 sites.}
\label{FCC_bond_one}
\end{figure}

Note that the jumps in the cluster sizes, as presented in Figs.~\ref{BCC} and
\ref{FCC}, are not infinitely sharp --- there is a slight rounding off. But this
is simply because these results are the {\it averages} over several
realizations, and the transitions occur at slightly different points in
different networks (a finite-size effect). But looking at each realization
individually, it turns out that in each case (and for both medium and large
networks) the transition usually happens in a matter of just {\it two bond additions}. In the
floppy phase, the largest  rigid cluster size is very small, usually around 10
(the maximum observed in the ten realizations for the largest size was 19) for
FCC and always 2 for BCC, and
there are never any stressed bonds; then all of a sudden, with a {\it single
bond} added, a huge cluster taking up more than 90\% of the network emerges; and
after just one more bond addition, a huge stressed region, again occupying
around 90\% of the network, appears.
Thus both the rigidity and the stress transitions
occur in the most dramatic manner possible, with
an enormous jump in the order parameter upon a single bond addition. This is
illustrated in Fig.~\ref{BCC_bond_one} for BCC and in Fig.~\ref{FCC_bond_one}
for FCC. In each case, the results for a single realization are shown. Note
that the interval between adjacent data points is {\it one bond}.

The width
of the rigidity transition in terms of $p$ thus appears to be ${\cal O}(1/N)$. Normally,
first order transitions are not as sharp, as different parts of the system
undergo transitions at slightly different points, resulting in the width that,
while decreasing with growing $N$, does so more slowly than ${\cal O}(1/N)$.
Another interesting point is the absence of the hysteresis problem commonly
associated with numerical studies of first order transitions: in thermal phase
transitions, for instance, as the temperature is changed to drive the system
across the transition, it takes a long time for the system to reach the new
phase and equilibrate and so the transition is delayed. With no equilibration
required and the pebble game being exact, there is no hysteresis here, of
course.

\begin{figure}
\begin{center}
\includegraphics[width=3in]{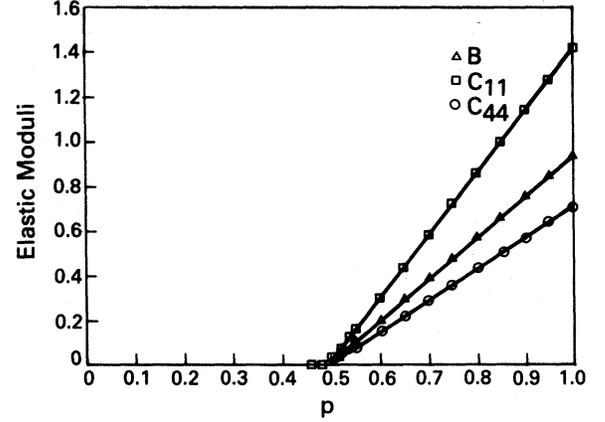}
\end{center}
\caption{The elastic moduli for the bond-diluted FCC network (adapted from
Ref.~\cite{garboczi}).}
\label{garboczi}
\end{figure}

We should also note that previous studies for bond-diluted FCC~\cite{garboczi}
and BCC~\cite{sahimi} and site-diluted FCC~\cite{garbsite} networks have
indicated that the elastic moduli change continuously at the rigidity
transition, without a jump (Fig.~\ref{garboczi}). This is expected to be
true for site-diluted BCC networks as well. We can
thus say that the transition is {\it geometrically first order but physically
second order}, as the geometric order parameters such as the size of the
percolating cluster jump at the transition, but the physical quantities such as
the elastic moduli do not.

\subsection{Site-diluted networks}
\label{site}

We now describe briefly the results of a similar study for {\it site-diluted}
networks. Some of the results are radically different from those for the
bond-diluted case.

In site-diluted networks, a certain number of sites are deleted with all
associated constraints, but all remaining sites retain all connections to other
remaining sites. One can still define the mean coordination $\ra$ as the mean
number of remaining neighbors of a site, averaged over all {\it remaining}
sites. In the Maxwell counting approximation, the rigidity transition is still
at $\ra=6$. If sites are deleted at random and the fraction of remaining sites
is $p$, then on average a fraction $p$ of neighbors of each present site remain,
so $\ra=zp$, where $z$ is the coordination number of the full lattice. This is
the same relation as Eq.~(\ref{rp}) for bond-diluted networks, except in that case
the relation was exact even for finite networks, whereas in the site-diluted case it
is exact only in the thermodynamic limit, but should
still be good for large enough networks. Based on Maxwell
counting, the transition should occur at $p\approx 1/2$ for FCC and
$p\approx 3/4$ for BCC, just like for bond dilution.

Similarly to the bond dilution case, we first use the relaxation algorithm. In
the BCC case, we consider 1000 realizations on site-diluted lattices, originally
of 686 sites each, all with 510 present sites ($p\approx 0.743$). At this point,
both percolating and non-percolating networks are present. There is no
qualitative difference with the case of bond dilution: still, only very small
(up to 3 sites) or very large (at least 459 sites) clusters are present. This is
to be expected: we know that clusters of sizes above 3 and below 84
cannot exist in principle, and this does not depend on the dilution
procedure, of course; non-percolating clusters of size 84 or larger, on the
other hand, are still expected to be very rare, although not as rare as in the
bond case, for reasons explained in the Appendix in the discussion of the FCC
case. Of course, no ``dangerous'' hinges, other than the trivial ones shared by
the percolating cluster and a triangle, are possible, so no problems in the
pebble game are expected, as far as the floppy mode count and the percolating
cluster size are concerned.

In the FCC case, on the other hand, the situation is very different. We consider
1000 realizations on site-diluted lattices, originally
of 500 sites each, all with 235 present sites ($p=0.47$). Both
percolating and non-percolating networks are present, as well as many
percolating in just one or two directions. In contrast to the
case of bond dilution, we now see many clusters of all sizes, not just very
small and very large. This is what one might expect in case of a
{\it second order} transition. Some statistics in comparison to the
case of bond dilution is given in the Appendix.

Given that many clusters of all sizes are present in the FCC site-diluted case,
one might naively expect many ``dangerous'' hinges with ``bananas'' on all
length scales. Fortunately, this turns out not to be the case. There are still
numerous hinges rigidifying triangles and tetrahedra that are also present in
the bond-diluted case (although now they are often shared with another
medium-sized cluster, not necessarily with the percolating cluster, which was
very rare in the bond-diluted case). Like before, the presence of these at most
means that the pebble game may fail to detect some triangles or tetrahedra ---
any larger clusters, including the percolating cluster, are not affected, nor is
the floppy mode count. Besides these, only very few other hinges are present. In
the same 1000 500-site networks, the relaxation algorithm detects 45 hinges
involving clusters both of which are larger than a tetrahedron and thus with
the potential to introduce more significant errors than missing a triangle or
a tetrahedron. Of these, 11 can influence the floppy mode count. But this is
still an overestimate, as we know, and needs to be confirmed by analyzing
stress. This more careful analysis brings the number of ``dangerous'' hinges
down to 18, or roughly one per 13000 present sites, of which {\it none} can
affect the floppy mode count. The largest affected cluster is still fairly
small, consisting of 24 sites, so the percolating cluster can never be
misidentified. Most misidentifications are, in fact, fairly benign --- 14 out of
18 involve missing a single triangle from a larger cluster and two more involve
missing a single tetrahedron. Given this, we can still safely use the pebble
game --- chances of the floppy mode count or the percolating cluster size being
affected are very slim.

\begin{figure}
\begin{center}
\includegraphics[width=3in]{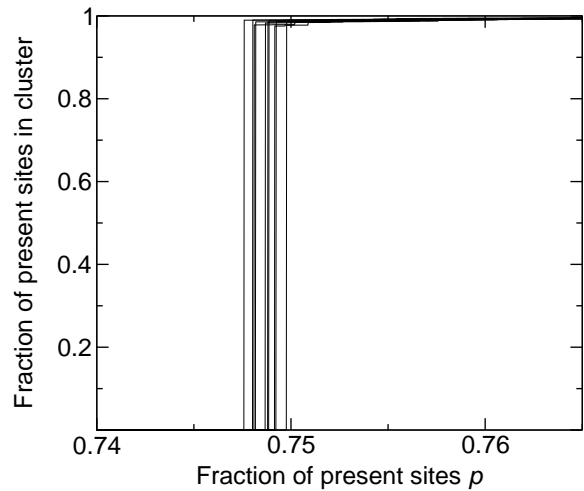}
\end{center}
\caption{The fraction of sites (among those still present, i.e., undeleted) in
the largest rigid cluster for the
site-diluted central-force BCC networks. Results for 10 realizations on networks
initially consisting of 54000 sites are plotted separately. The step between
adjacent data points is one site.}
\label{BCC_site}
\end{figure}

\begin{figure}
\begin{center}
\includegraphics[width=3in]{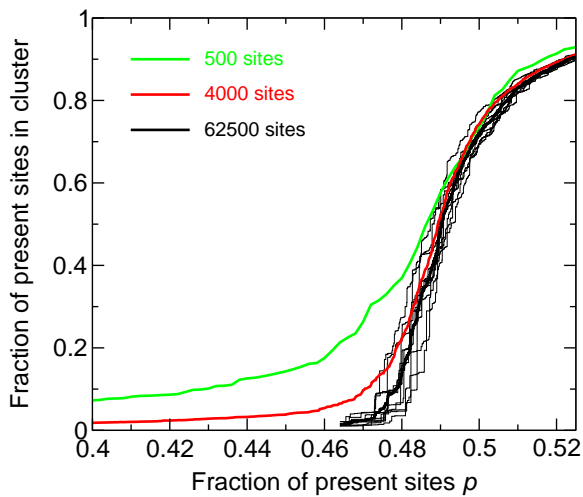}
\end{center}
\caption{The fraction of sites (among those still present, i.e., undeleted) in
the largest rigid cluster for the site-diluted central-force FCC networks. For
the smallest (500 sites initially) and medium (4000 sites) networks, the
average over 100 realizations is plotted; for the largest (62500 sites)
networks, both the average over 10 realizations (the thick black line) and
the results for the individual realizations (thin lines) are plotted. The step
between adjacent data points is one site in all cases.
}
\label{FCC_site}
\end{figure}

The pebble game results confirm the expectations based on the relaxation study.
The largest  rigid cluster size as a function of the fraction of sites
present is shown in Fig.~\ref{BCC_site} for the BCC lattice and in
Fig.~\ref{FCC_site} for the FCC lattice. In the BCC case, we again see very
sharp jumps, with the
largest cluster size changing from 2 to almost all network upon addition of a
single site, consistent with a first order transition. But in the FCC case,
the largest  cluster size grows gradually when sites are added, with only very
small jumps present in individual realizations, which is typical of a second
order transition.

\subsection{Discussion}
Systems that we have considered in this section differ just in the underlying
lattice and/or type of disorder (site vs. bond). It is usually thought that
these properties are irrelevant when determining general properties of a phase
transition, such as its order, or, in the case of a second order transition,
the critical exponents, unless long-range interactions are introduced. For
instance, in the Potts model~\cite{wu}, the order of the transition
depends on the dimensionality and the number of states, but not on the lattice
type. It is certainly possible to have a {\it tricritical point} separating
regions of first and second order transitions, but this is usually observed when
there are several competing interactions whose relative strengths can be varied.
This is not the case here, and thus the situation we observe is highly
unusual and counter to our intuitive expectations based on universality.

Of course, it should be remembered that claims based on numerical simulations
can rarely be made with absolute certainty. While our results in
Fig.~\ref{FCC_site} give a strong indication that the rigidity transition is
second order in the case of FCC site dilution, a first order transition
rounded due to finite-size effects can never be ruled out completely. Our claim
that in the other three cases the transition is first
order looks even stronger given how sharp the transition is; but even in this
case, surprises are possible.

\section{A counterexample: chains of edge-sharing tetrahedra}
\label{edgesharing}

In the previous section, we have shown that in the particular case of randomly
diluted central-force networks, there are virtually no errors in the pebble game, except
insignificant ones, such as missing a small cluster. In particular, there are
hardly any double-banana-type configurations similar to those shown in
Figs.~\ref{2banana} and \ref{8plus8}. However, such configurations may be quite
frequent in certain cases.

\begin{figure}
\begin{center}
\includegraphics[width=3in]{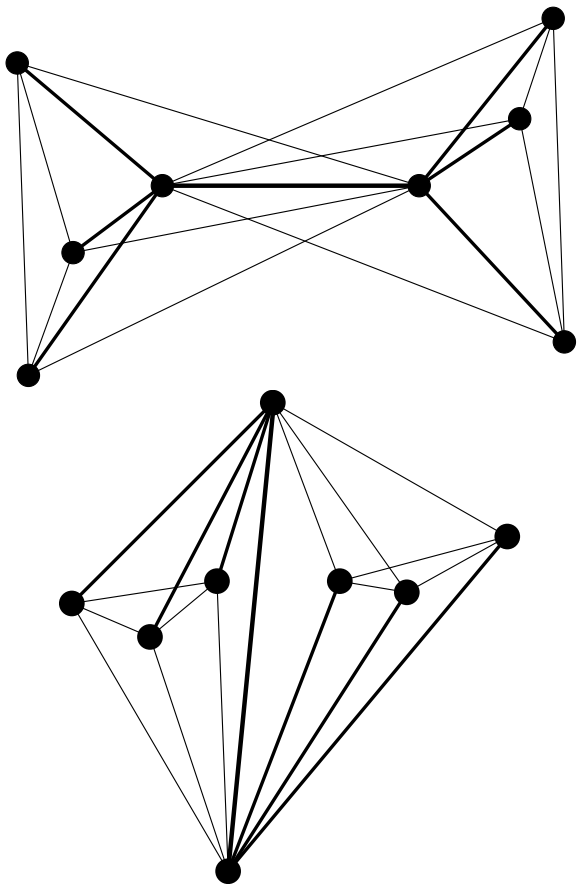}
\end{center}
\caption{At the top, two connected 4-fold coordinated sites and
their neighbors in a bond-bending network with both
central-force (thick lines) and bond-bending (thin lines) constraints
shown. This is equivalent to a double-banana graph with the hinge added
explicitly (shown at the bottom).}
\label{2tetra}
\end{figure}

For instance, consider
two neighboring 4-fold coordinated sites in a bond-bending network. It is
easy to realize that a network consisting of such a pair of sites with their
neighbors and associated central-force and bond-bending constraints
is topologically equivalent to the double-banana graph plus the explicit
hinge (Fig.~\ref{2tetra}). Specifically, each banana is formed by the constraints
associated with a particular site of the pair, and the constraint connecting these
sites is the hinge. If one inserts constraints in an arbitrary order during the
pebble game, it is possible that the hinge is inserted last (after all other
central-force and angular constraints), in which case it will not be covered by
a pebble and according to the analysis in Section~\ref{errors}, will cause an error in
the pebble game floppy mode count, as well as rigid cluster decomposition (just
as if the hinge were implicit). This is why it is
important to insert constraints in a proper order for the pebble game to be
correct for bond-bending networks.

\begin{figure}
\begin{center}
\includegraphics[width=3in]{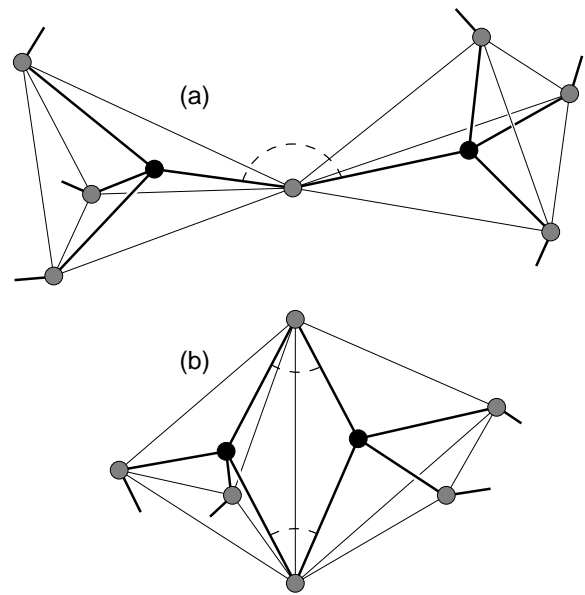}
\end{center}
\caption{Fragments of networks consisting of atoms of valence 2 and 4.
(a) A pair of corner-sharing tetrahedra. (b) A pair of edge-sharing tetrahedra.
In both cases, thick lines are first-neighbor constraints and thin lines are
second-neighbor (angular) constraints; black atoms
have valence 4 and have
all associated angular constraints present, while gray atoms
have valence 2
and their angular constraints are missing, so that angles marked with
dashed arcs are {\it not} constrained. It is implied that these
pairs of tetrahedra are connected to the rest of the network, as shown by short
black lines stemming from atoms of valence 2. In particular, in the edge-sharing
case, each of the two tetrahedra can share an edge with yet another tetrahedron,
and thus a chain of edge-sharing tetrahedra will be formed.}
\label{tetra}
\end{figure}

Another spectacular case is a chain of {\it edge-sharing tetrahedra}. Consider
a covalent network consisting of atoms of valence 4 (such as Si or Ge)
and 2 (Se, Te, S, or O). Suppose there is perfect chemical order, i.e., each
atom of valence 4 is always next to an atom of valence 2 and vice versa. Suppose
also that there are both central-force and angular constraints associated with
atoms of valence 4. Then an
atom of valence 4 with all its associated
constraints forms a rigid object that can be thought of as a rigid tetrahedron,
where the atom itself is at its center and the 4 neighbors (all of which are
atoms of valence 2) are the vertices. Then the network can be represented
as a system of connected tetrahedra. These tetrahedra can be
corner-sharing [Fig.~\ref{tetra} (a)] or edge-sharing [Fig.~\ref{tetra} (b)].
Suppose now that
angular constraints at atoms of valence 2 are
weaker and can be neglected
in a crude approximation (this is often the case when these are oxygen
atoms~\cite{zeolites}). Then in the edge-sharing case one ends up with chains of
edge-sharing tetrahedra that can rotate with respect to each other around a
common edge. Two tetrahedra with a common edge are topologically equivalent to a
double-banana graph, with the common edge being the hinge. So a chain of
edge-sharing tetrahedra is actually a chain of bananas. There are as many floppy modes
(in addition to rigid body motions of the whole chain) as there are hinges (or one less than
there are tetrahedra in the chain). In the worst-case scenario, about a half
of all hinges may end up being uncovered during the pebble game, and then the
pebble game will miss about a half of the floppy modes. Note that here the hinges are
{\it second} neighbor constraints, so it is safer to insert second-neighbor
constraints first to make sure they are covered --- the situation
opposite to that with purely bond-bending
networks. Of course, the worst-case scenario is unlikely --- in reality,
in the case of random insertion, it can be estimated numerically that in long
chains, about 13.5\% of the floppy modes are going to be missed.

\section{Conclusion}
In this paper, we have described an extension of the pebble game algorithm
for rigidity analysis that was used previously for the special class of
bond-bending networks in 3D. The new algorithm is applicable to general 3D
networks, but is
approximate: there are networks for which there are errors in the number of
floppy modes, rigid cluster decomposition and/or finding stress. We have also
introduced a slower but exact algorithm, the relaxation algorithm.
Unlike the pebble game, it is not an integer algorithm (it involves floating
point operations), but it has a number of built-in consistency
checks, so
errors due to rounding are unlikely in the final result. The relaxation
algorithm can be used for comparison with the pebble game using a few
representatives of a particular class of networks, before the latter algorithm
is used more extensively.

Using the relaxation algorithm and other considerations, we have argued that
for randomly diluted central-force networks, the pebble game algorithm is
essentially exact, as far as the percolating cluster size, stressed
bonds, and the number of floppy modes are concerned; errors are possible, but
extremely rare. Applying the pebble game to {\it bond-diluted} networks, we
conclude that the rigidity percolation transition on such
networks is first order, in contrast to bond-bending networks in 3D and
central-force networks in 2D. In fact, the transition is actually first order
geometrically, but second order physically, as it is known from previous work
that the elastic constants change continuously at the transition. On the other
hand, for {\it site-diluted} networks, the order of the transition
depends on
the lattice type: first order for BCC and second order for FCC. The dependence
of the order of the transition on the lattice type and the disorder type would be highly
unusual and even though the evidence we present is rather strong, further
research is needed to confirm our results.

At the same time, there are networks for which the pebble game is less
successful. In particular, this is so for bond-bending networks, when the proper
order of constraint insertion is not obeyed, and partially bond-bending
networks, one example of which, chains of edge-sharing tetrahedra, is discussed
in the paper. What makes the difference between the ``good'' and the ``bad'' networks?
In randomly diluted central-force networks, medium-sized and large clusters are
relatively rare, even in site-diluted FCC nets in which they are much more
frequent than in the other three cases considered here --- much of the network
is in clusters of sizes below 4 or in the percolating cluster. It is even more
rare for two of such medium or large clusters to touch at exactly two places and
form a hinge. But in bond-bending and partially bond-bending networks, every
site of coordination at least 4 with its
angular constraints present is associated with a cluster of at least site 5,
moreover, the vicinity of such a site is always stressed. If at the same time the
{\it average} coordination of the network is low so the network is floppy
overall, there will be many implied or stressed explicit hinges between such
medium-sized clusters that can lead to errors in the pebble game. Arguably,
such situations are less frequent than those in which such hinges are rare, but
in each case tests should be run before the pebble game algorithm is used.

Of course, ideally one would desire an {\it exact} integer algorithm for
rigidity analysis. Efforts to design such an algorithm have
not paid off so far, and the authors would argue based on this experience that
developing a polynomial-time pebble-game-type algorithm, while very
interesting for the computer science and mathematical rigidity theory community,
would be useless in practice: it is unlikely that such an algorithm would be
sufficiently fast to beat the relaxation algorithm which is already appropriate
for many purposes.

\section*{Acknowledgments}
We should like to thank D.J.~Jacobs and W.~Whiteley for many discussions and
insights over the past 4 years that have contributed many ideas that helped lead
to the work described in this paper. We have also benefited from discussions
at the Workshop on Modeling Protein Flexibility and Motions in Banff, Alberta,
Canada (July 2004). We also acknowledge support from the NSF under grant DMR-0425970 and
the Natural Sciences and Engineering Research Council of Canada (NSERC), and
thank the R\'eseau qu\'eb\'ecois de calcul de haute performance (RQCHP) for
computer resources.

\section*{Appendix: small clusters in the central-force diluted BCC and FCC
networks}
Here we look in more detail at possible small rigid clusters and stressed
regions in central-force diluted BCC and FCC networks. As we have mentioned, the reason the
rigidity transition is first order in the previously considered case of
RBNs has to do with the absence of finite rings and thus
finite rigid clusters in these networks. While the same clearly cannot be true
for regular BCC and FCC lattices, it is interesting to find out if there are
any similarities.

We first find what cluster sizes are possible. For this, we use a
computational procedure outlined below. Its advantage compared to just looking
at rigid cluster decomposition of many networks is that even if clusters of
certain size are extremely rare and would never be seen in reality, they can
still be found with our approach. The problem of generating rigid clusters
resembles that of generating lattice animals in ordinary, connectivity
percolation~\cite{stauffer}.

Basically, the idea of the approach is to generate the most rigid configuration
of a given size through an optimization procedure. First, note that a rigid cluster
with $n$ sites should contain at least $3n-6$ constraints (if it is rigid by itself), or
$3n-7$ constraints, if it shares a hinge with some other rigid cluster and is
rigidified by it. So, if rigid clusters of size $n$ are possible, the maximum
possible number of constraints (where the maximum is taken over all possible
graphs with $n$ sites on the full lattice) should be at least $3n-6$ (or
respectively $3n-7$). Then the purpose is to
maximize the number of constraints (always equal to the number of bonds in
central-force networks) for given $n$. This is done as follows. First of
all, note that if the positions of the $n$ sites on
the lattice are fixed, the optimum bond configuration consistent with those fixed
positions will be the one containing all possible bonds connecting the sites
in these fixed positions. Then the idea is to move sites around on the lattice
and while doing so try to maximize the number of possible connections
between those sites. This can be done using an analog of the simulated annealing
procedure~\cite{siman}, where the "energy" that needs to be minimized is the negative number
of connections for the given configuration of sites (as this is the quantity
being optimized). If the optimization is done carefully, we can be sure that the
optimum number of bonds for given $n$ is reached. If this optimum number is
less than $3n-7$, then a rigid cluster with $n$ sites is not possible (we
neglect the possibility of a cluster being rigidified by more than one
external rigid cluster: at least in the bond-diluted case, rigid clusters of
medium and large size, other than the single percolating cluster, are rare, so
two of them coming together to rigidify a third one is even more rare; even in
the FCC site case, when medium-sized clusters are not rare, we have not seen
such a situation). If the
optimum number is exactly $3n-7$, then a standalone rigid cluster of size $n$
is not possible, but a cluster rigidified from the outside (say, by the
percolating cluster) and sharing a hinge with it (a situation
similar to that in Fig.~\ref{2banana1}) is possible. If the optimum number of bonds is
exactly $3n-6$, then a standalone rigid cluster is possible, but any such
cluster will be isostatic (stress is impossible). Finally if the optimum
number is $3n-5$ or higher, then even a stressed region of size $n$ is possible.
One caveat is that even a network with, say, $3n-6$ bonds may be floppy, if it
contains a stressed region. Such situations are, however, easy to detect: start
with the smallest $n$, find the maximum number of bonds, then go to
$n+1$, etc.; if the difference between the maximum number of bonds and the
number of degrees of freedom ($=3n$) ever goes {\it down} with increasing $n$,
this means that the network actually became floppier with increasing size and
rigid clusters of the last analyzed size are impossible, even if the number of
constraints is still $3n-6$ or more.

Using the above procedure, we have found possible sizes of rigid clusters and
stressed regions for both FCC and BCC networks. For BCC, clusters of size 2
(single bonds) are, of course, possible, but standalone clusters of size 3
are not, because the BCC lattice does not contain triangles. However, clusters
of size 3 rigidified by another cluster are possible (indeed, we have observed
many clusters of size 3 sharing a ``trivial'' implied hinge with the percolating
cluster, as we have discussed in subsection~\ref{relaxCF}). Clusters of sizes
4, 5, etc. (either standalone or not) are {\it not} possible, however; the
next possible non-standalone cluster size is 84 and the next possible standalone
cluster size is 90. The smallest possible standalone stressed region has size
96. So even though, unlike in RBNs, rings of small sizes are possible, small
rigid clusters (other than trivial single sites, single bonds and triangles) are still not
possible, like in RBNs. Of course, in RBNs clusters of {\it any} finite size are
not possible, and here clusters of size 84 and higher can exist at least in
principle. But in practice, such clusters are extremely rare, as we show below.

For FCC, the situation is different. It turns out that rigid clusters of
{\it all} sizes are possible (except standalone clusters of size 5 are not
possible, but non-standalone clusters of this size are still possible).
However, using the relaxation
algorithm, we have seen that the frequency of clusters decreases very fast with
their size: there are a lot of clusters of size 2 and 3, clusters of size 4 are
much more rare and there are almost no clusters of a larger size.
Standalone stressed regions, on the other hand,
start at size 10 and these are much more rare (non-standalone stressed
regions are essentially double-banana situations, and we discuss them
separately, in subsection~\ref{relaxCF} and later in this Appendix).

Obviously, the optimization procedure described above not only answers the
question about the possibility of a cluster of a certain size, but also
produces a realization of such a cluster when the answer is positive. If the
optimization procedure is carried out many times starting from different
initial configurations and using different random number sequences in the
simulated annealing, eventually the set of final configurations will reproduce
the complete set of possible rigid clusters. This allows an estimate of
probabilities of certain clusters.

We consider the case of bond dilution first
and then make a comparison to site dilution at the end of this Appendix.
For simplicity, we will consider the situation when all clusters of a given size
$n$ have the same number of bonds $n_B$ equal to the maximum possible one. In
FCC, for instance, this will be the case for all clusters of sizes
smaller than 10 (the smallest stressed region). We also consider standalone
clusters only at this point. Suppose the total number per site of possible
rigid cluster configurations of size $n$ is $c_n$ (this includes all configurations related
by symmetry). Then the frequency per site of clusters of size $n$,
$\nu_n$, can be estimated as
\begin{equation}
\nu_n\approx c_n p^{n_B},\label{freq_bond}
\end{equation}
where $p$ is the fraction of
present bonds in the network. This
involves three assumptions: (a) such clusters are
rare so ``excluded volume'' effects can be neglected; (b) the
cluster frequency decreases fast with increasing cluster size, so that the
probability of clusters of size larger than $n$ can be neglected when
calculating the probability of clusters of size $n$; (c) there is no
percolating cluster.

Table~\ref{table1} gives the frequencies observed in simulations for FCC
networks using both the relaxation algorithm and the pebble game, compared to
the estimates obtained using Eq.~(\ref{freq_bond}) with $n_c$ calculated using
the ``simulated annealing'' procedure described above and assuming $n_B=3n-6$,
which is true for $n<10$ (for this reason, we do not give estimates for $n=10$
and above). For the simulation results, we use just non-percolating networks,
since Eq.~(\ref{freq_bond}) ignores the possibility of existence of the percolating cluster. We
also note that nearly all clusters in non-percolating networks are standalone,
which is another assumption behind Eq.~(\ref{freq_bond}). For the relaxation
study, we use the sample of 3100 500-site networks described
in subsection~\ref{relaxCF}. Of these, there are 1812 non-percolating networks,
with a total of 906 000 sites. Thus, to obtain the frequency per site, the
total number of observed clusters of a given size is divided by 906 000. For the
pebble game study, we also use 500-site networks, but a much higher number of
them: $10^7$ networks, of which more than $9\times 10^6$ are non-percolating
(a much higher percentage of non-percolating networks compared to the sample
used in the relaxation study is simply due to the fact than in the pebble game
runs, we fix the bond number to 1470, which is slightly below the transition,
whereas in the relaxation runs, the number of bonds varies between 1460 and 1490
and this spans the whole transition region). Much better statistics we are able
to obtain with the pebble game illustrates its advantage compared to the
relaxation algorithm. At the same time, it can be seen
that the frequencies obtained using these two methods are essentially the same,
apart from the statistical noise; this is yet another confirmation of the
accuracy of the pebble game algorithm. For the theoretical estimate, we use
$p=0.49$; this corresponds to
1470 bonds, which is the same number used in the pebble game runs and close to
the average over the relaxation runs. It is seen that the frequency tends to
be overestimated somewhat, especially for larger sizes, so rare
larger clusters, hinges, etc., may even be more rare than the estimates
below suggest.

\begin{table*}
\caption{Frequencies per site of different cluster sizes in 500-site
bond-diluted FCC networks obtained both with the relaxation algorithm (using
the 1812 non-percolating networks out of a sample of 3100 networks with
numbers of bonds between 1460 and 1490) and with the pebble game
(using the 9242577 non-percolating networks out of a sample of $10^7$ networks
with 1470 bonds). This is compared to the theoretical estimates obtained using
Eq.~(\ref{freq_bond}), with quantities $c_n$ obtained using the numerical
``simulated annealing'' procedure described in the text, and with the bond
number $n_B=3n-6$.}
\begin{tabular}{|r||r|r||r|r||r|r|}
\hline
cluster  & \multicolumn{2}{c||}{relaxation} & \multicolumn{2}{c||}{pebble game} &       & Theoretical \\
\cline{2-5}
size $n$ & \# clusters & frequency         & \# clusters & frequency          & $c_n$ & frequency   \\  
\hline\hline
3        & 747196      & $8.2\times 10^{-1}$ & 3824801279 & $8.3\times 10^{-1}$ & 8  & $9.4\times 10^{-1}$ \\
4        & 24665       & $2.7\times 10^{-2}$ & 125596870  & $2.7\times 10^{-2}$ & 2  & $2.8\times 10^{-2}$ \\
5        & 0             & 0                  & 0          & 0                    & 0  & 0                    \\
6        & 67          & $7.4\times 10^{-5}$  & 318927     & $6.9\times 10^{-5}$ & 1  & $1.9\times 10^{-4}$ \\
7        & 78          & $8.6\times 10^{-5}$  & 329476     & $7.1\times 10^{-5}$ & 8  & $1.8\times 10^{-4}$ \\
8        & 26          & $2.9\times 10^{-5}$  & 134915     & $2.9\times 10^{-5}$ & 28 & $7.4\times 10^{-5}$ \\
9        & 5           & $5.5\times 10^{-6}$  & 36639      & $7.9\times 10^{-6}$ & 80 & $2.5\times 10^{-5}$ \\
10       & 1           & $1\times 10^{-6}$    & 13145      & $2.8\times 10^{-6}$ &    &                      \\
11       & 1           & $1\times 10^{-6}$    & 7334       & $1.6\times 10^{-6}$ &    &                      \\
\hline
\end{tabular}
\label{table1}
\end{table*}

We can now use our procedure to estimate the frequencies of finite clusters
of size larger than 3 in BCC networks. We use $p=0.745$. The smallest standalone
cluster has size 90 and $3\times 90-6=264$ bonds; the optimization procedure
gives $c_{90}=27$. Then
$\nu_{90}\approx 27\times 0.745^{264}\approx 5\times 10^{-33}$. The next
possible size is 94, with 276 bonds; $c_{94}=720$; then
$\nu_{94}\approx 4\times 10^{-33}$. These probabilities are extremely small;
obviously, existence of finite clusters of size larger than 3 can be neglected
for any practical purposes, and in this respect, BCC networks are exactly
like RBNs, even though they are ``normal'', regular networks with loops.

A similar procedure can be used to find possible double-banana-type networks.
For this, a rigid cluster is chosen and fixed to serve as the first
``banana''; two sites belonging to it are also fixed as hinge ends. A set of
sites is then allowed to move around on the lattice, like in the previous case,
but with the restriction that no sites can ever coincide with any of the sites
belonging to the first ``banana'', {\it except} for two of them that should
{\it always} coincide with the assigned hinge ends. This should be tried for
all possible ``first bananas'' and for each ``banana'', for every possible pair of
hinge ends. The result of this procedure is that in FCC networks, the
smallest double-banana-type graph with an implied hinge that will violate the
floppy mode count when the pebble game is run consists of two 8-site
``bananas'', as shown in Fig.~\ref{8plus8}. We have already mentioned in
subsection~\ref{relaxCF} that these are very rare --- probably less than one in a
billion sites. On the other hand, the smallest
two-cluster graph with a hinge that cannot influence the correctness of the
floppy mode count, but can influence the correctness of rigid cluster
decomposition, consists of an 8-site cluster and a triangle. These should be
more frequent, although we have not seen them in our relaxation
simulations, even though we have seen
26 8-site clusters; there were a few hinges of this type involving the
10- and 11-site clusters (each seen once in our relaxation simulations).

So far, we have mostly concentrated on the floppy phase. We now look at the
small clusters in the rigid phase, again for obvious reasons concentrating on
FCC networks. In the rigid phase, the percolating cluster takes up most of the
network. Small clusters larger than triangles are rather rare. In all 1288
percolating networks, only 14 4-site clusters were observed, and larger
clusters were never seen. Part of the reason is obviously the fact that only
a small part of each network is not in the percolating cluster. Perhaps more
importantly, for a small rigid cluster to exist, not only should there be
enough bonds locally for the configuration to be rigid (which was enough in
the floppy phase), but also this
configuration should be only sparsely connected to the rest of the network, or
else it will be part of the percolating cluster instead. At most, a small
rigid cluster can touch the percolating cluster in two points (thus sharing
a hinge); it is also possible that besides this, it is connected to the
percolating cluster via a chain with at least two links (then the cluster will
still remain floppy with respect to the percolating cluster), but such
chains are also rather rare. So the conditions of existence for larger
non-percolating clusters are rather stringent and become progressively more
stringent as the size increases, since
it becomes harder and harder to ensure that the region is sufficiently
disconnected from the rest of the network.

For example, the smallest graph of the double-banana type in the rigid phase is
the combination of a percolating cluster and a 6-site cluster (an octahedron),
with the hinge between two opposite vertices of the octahedron. Octahedra
are rare in the floppy phase, but not totally unseen (there is roughly one
per 10000 sites, according to Table~\ref{table1}). But this should not be a
reason for concern: octahedra, especially those sharing an implied hinge with the
percolating cluster, are much more rare in the rigid phase, for reasons explained
above. As mentioned, we have not seen them in our relaxation simulations.
In a much larger pebble game run, with $3\times 10^7$ networks of 1372 sites
each, and about 20\% of networks percolating, there were 21 octahedra, or about
one per $4\times 10^8$ sites. Given that only a small fraction of these are
expected to share an implied hinge with the percolating cluster, the frequency
of implied hinges of this type is certainly less than $10^{-9}$ and probably
much less than that.

In the case of site dilution, on the other hand, the number of medium-sized
clusters is much higher. Table~\ref{table2} shows the frequencies for the same
cluster sizes as in Table~\ref{table1} obtained using the relaxation
algorithm in 1000 realizations on
500-site FCC lattices, all with the fraction of present sites $p=0.47$ (235
sites), slightly below the transition (see Fig.~\ref{FCC_site}). The frequency
is given per {\it present} site; unlike in Table~\ref{table1}, both percolating
and non-percolating networks were used (averages over non-percolating networks
only would likely be even higher). While the frequencies of 6-site clusters are
about the same in the bond and site cases, 8-site clusters are about 50 times
more frequent in the site case, and the ratio reaches about 300 by
$n=10$. Even though analytical estimates of the type
done for bond dilution are much harder to make in the site case, since the
assumption about the cluster frequency decreasing rapidly with the size is no
longer valid, qualitatively it is clear what makes the difference between the
two cases. Assume that a certain set of $n$ sites can be mutually rigid in
principle in the FCC network. In the site dilution case, for this set to
actually be rigid, these $n$ sites just need to be present, since all
connections between them are guaranteed automatically --- the probability of
this is $p^n$. But in the case of bond dilution, bonds can be present or absent
independently. At least $3n-6$ of them need to be present, and for $n<10$, this
is also the maximum total number of bonds that a cluster of size $n$ can have.
Then the probability of this is $p^{3n-6}$. Given that the transition in the
site and bond cases occurs at about the same value of $p\approx 1/2$, it is
clear that clusters of the same size are more frequent in the site case. These
arguments are clearly very simplistic: for the cluster to be of exactly size
$n$, not only should the $n$ sites be mutually rigid, but no other sites should
be rigid with respect to them all (taking this into account is especially
important in the site case); also the arguments need to be modified for
$n\ge 10$, when the maximum possible number of bonds is not $3n-6$. But the
main idea remains valid.

\begin{table}
\caption{Frequencies per present site of different cluster sizes in site-diluted
FCC networks, obtained using the relaxation algorithm in 1000
realizations on 500-site lattices with 235 present sites.}
\begin{tabular}{|r|r|r|}
\hline
Cluster  & Observed    & Observed    \\
size $n$ & \# clusters & frequency   \\
\hline\hline
3   & 159147      & $6.8\times 10^{-1}$ \\
4   & 20881       & $8.9\times 10^{-2}$ \\
5   & 1           & $4\times 10^{-6}$    \\
6   & 23          & $9.8\times 10^{-5}$  \\
7   & 202         & $8.6\times 10^{-4}$  \\
8   & 375         & $1.6\times 10^{-3}$ \\
9   & 273         & $1.2\times 10^{-3}$ \\
10  & 197         & $8.4\times 10^{-4}$  \\
11  & 161         & $6.9\times 10^{-4}$  \\
\hline
\end{tabular}
\label{table2}
\end{table}

\end{document}